\documentclass[aps,prx,twocolumn,footinbib,superscriptaddress]{revtex4}
\usepackage{graphicx}% 
\usepackage{siunitx}
\usepackage{bm}% bold math
\usepackage[caption=false]{subfig}
\usepackage{mathtools}
\usepackage{multirow}
\usepackage{amssymb}
\usepackage{braket}
\usepackage{amsmath}
\usepackage{placeins}
\usepackage{cancel}
\usepackage{xcolor}
\usepackage{spreadtab}
\usepackage{dsfont}
\usepackage[colorlinks=true,
            linkcolor=blue,
	        citecolor=blue,
	         urlcolor=blue]{hyperref}

\newcommand\bk{{\mathbf{k}}}
\newcommand\bq{{\mathbf{q}}}

\renewcommand{\arraystretch}{1.2}

\begin{document}

\title[Article Title]{Charting the landscape of Bardeen--Cooper--Schrieffer superconductors in experimentally known compounds}

\author{Marnik Bercx}
\email{marnik.bercx@psi.ch}
\affiliation{%
PSI Center for Scientific Computing, Theory, and Data, and National Centre for Computational Design and Discovery of Novel Materials (MARVEL), 5232 Villigen PSI, Switzerland.
}%
\author{Samuel Ponc\'e}
\email{samuel.ponce@uclouvain.be}
\affiliation{%
European Theoretical Spectroscopy Facility, Institute of Condensed Matter and Nanosciences, Université catholique de Louvain, Chemin des Étoiles 8, B-1348 Louvain-la-Neuve, Belgium. 	
}%
\affiliation{% 
WEL Research Institute, avenue Pasteur, 6, 1300 Wavre, Belgique.		
}%
\author{Yiming Zhang}
\affiliation{%
European Theoretical Spectroscopy Facility, Institute of Condensed Matter and Nanosciences, Université catholique de Louvain, Chemin des Étoiles 8, B-1348 Louvain-la-Neuve, Belgium. 	
}%
\author{Giovanni Trezza}
\affiliation{%
Department of Energy, Politecnico di Torino, Italy. 	
}%
\author{Amir Ghorbani Ghezeljehmeidan}
\affiliation{%
Electronic Components, Technology and Materials (ECTM), TU Delft, The Netherlands. 	
}%
\author{Lorenzo Bastonero}
\affiliation{%
Bremen Center for Computational Materials Science, and MAPEX Center for Materials and Processes, University of Bremen, 28359 Bremen, Germany. 	
}%
\author{Junfeng Qiao}
\affiliation{%
Theory and Simulation of Materials (THEOS), and National Centre for Computational Design and Discovery of Novel Materials (MARVEL), \'Ecole Polytechnique F\'ed\'erale de Lausanne (EPFL), 1015 Lausanne, Switzerland. 	
}%
\author{Fabian O. von Rohr}
\affiliation{%
Department of Quantum Matter Physics, University of Geneva, 24 Quai Ernest-Ansermet, 1211 Geneva, Switzerland. 	
}%
\author{Giovanni Pizzi}
\affiliation{%
PSI Center for Scientific Computing, Theory, and Data, and National Centre for Computational Design and Discovery of Novel Materials (MARVEL), 5232 Villigen PSI, Switzerland. 	
}%
\author{Eliodoro Chiavazzo}
\affiliation{%
Department of Energy, Politecnico di Torino, Italy. 	
}%
\author{Nicola Marzari}
\affiliation{%
PSI Center for Scientific Computing, Theory, and Data, and National Centre for Computational Design and Discovery of Novel Materials (MARVEL), 5232 Villigen PSI, Switzerland. 	
}%
\affiliation{%
Bremen Center for Computational Materials Science, and MAPEX Center for Materials and Processes, University of Bremen, 28359 Bremen, Germany. 	
}%
\affiliation{%
Theory and Simulation of Materials (THEOS), and National Centre for Computational Design and Discovery of Novel Materials (MARVEL), \'Ecole Polytechnique F\'ed\'erale de Lausanne (EPFL), 1015 Lausanne, Switzerland. 	
}%

\date{\today}

\begin{abstract}
We perform a high-throughput computational search for novel phonon-mediated superconductors, starting from the Materials Cloud 3-dimensional structure database of experimentally known inorganic stoichiometric compounds.
We first compute the Allen-Dynes critical temperature (T$_{\rm c}$) for 4533 non-magnetic metals using a direct and progressively finer sampling of the electron-phonon couplings. 
For the candidates with the largest T$_{\rm c}$, we use automated Wannierizations and electron-phonon interpolations to obtain a high-quality dataset for the most promising 250 dynamically stable structures, for which we calculate spectral functions, superconducting bandgaps, and isotropic Migdal-Eliashberg critical temperatures.
For 140 of these, we also provide anisotropic Migdal-Eliashberg superconducting gaps and critical temperatures. 
The approach is remarkably successful in finding known superconductors, and we find 24 unknown ones with a predicted anisotropic T$_{\rm c}$ above 10~\si{\kelvin}. 
Among them, we identify a possible double gap superconductor (p-doped BaB$_2$), a non-magnetic half-Heusler ZrRuSb, and the perovskite TaRu$_3$C, all exhibiting significant T$_{\rm c}$. 
Finally, we introduce a sensitivity analysis to estimate the robustness of the predictions.  
\end{abstract}

\maketitle

\section{Introduction}\label{sec1}

Superconductors play an important role in many modern technologies, finding applications in magnetic resonance imaging machines, maglev trains, and large-scale research infrastructures such as the large-hadron collider at CERN.
Over the past century, many classes of materials have been identified as new potential superconductors, but the most commonly used ones are still Nb-Ti alloys and A15 phases such as~Nb$_3$Sn~\cite{Uglietti2019} which are well described by the Bardeen-Cooper-Schrieffer (BCS) theory of superconductivity~\cite{Boeri2018,Boeri2019}.
High-temperature superconductors at ambient pressure such as cuprates~\cite{Shen2008} and iron-based~\cite{Fernandes2014} superconductors have much higher transition temperatures~($T{\rm _c}$) but often come with challenges that hinder their practical use, such as brittleness, anisotropic superconductivity requiring precise grain alignment, and low critical currents, not to mention a still elusive theoretical foundation. 
Until the last century, around one hundred stoichiometric, ambient pressure, BCS superconductors had been found~\cite{Hosono2015}, representing only one (on average) new superconductor every year since the discovery of superconductivity, with most recent efforts delivering near room temperature critical temperatures, albeit at ultra-high pressures~\cite{Zhang2017}. 
Despite extensive efforts, targeted or serendipitous discoveries have not fulfilled the demand for high-performance superconductors for industrial and scientific applications, and a cost-effective computational search is also desirable.

During the past few decades, first-principles calculations have played an increasingly important role in both understanding and predicting the material-specific aspects of superconductivity; see, e.g., Ref.~\cite{Boeri2018} for a review.
Although understanding high-temperature superconductivity is still challenging~\cite{OMahony2022}, conventional BCS superconductors are typically tractable and can be studied readily via well-established techniques~\cite{Baroni2001,Giustino2017,Mori2024}.
Therefore, a number of efforts have recently emerged to find new phonon-driven superconductors using ab-initio evolutionary search~\cite{Kolmogorov2010,Gou2013}, machine learning approaches~\cite{Stanev2018,Xie2022,Choudhary2022,Tran2023,Wines2023a,Pogue2023,li2024,han2024,Lee2024} or high-throughput screening~\cite{Shipley2021,Choudhary2022,Cerqueira2023,Wines2023,Tran2024}.
Machine learning approaches are often (but not exclusively~\cite{Sommer2023}) based on training on the experimental SuperCon database~\cite{Hosono2015,li2024}, which contains chemical formulae but not the crystal structures and focuses primarily on non-conventional superconductors. 
Instead, high-throughput studies based on first-principles calculations have relied on coarse momentum grid integrations for screening and  performed accurate calculations on a small number of promising candidates. 
As shown before~\cite{Margine2013} and also highlighted here, a precise calculation of the electron-phonon interactions, essential for determining superconductivity, typically requires an ultra-dense sampling of the Brillouin zone.

The \textsc{EPW} (Electron-phonon using Wannier functions) code~\cite{Ponce2016, Lee2023} can perform such precise interpolations of the electron-phonon matrix elements at a low computational cost.
However, performing an extensive search using these advanced methods remained elusive until now, as obtaining the relevant maximally-localized Wannier functions (MLWFs)~\cite{Marzari2012} historically required a combination of chemical intuition and trial-and-error efforts.
In this work, we combine recent developments in automated Wannierization algorithms~\cite{Damle2018} with the \textsc{AiiDA} computational infrastructure~\cite{Huber2020,Uhrin2021} to perform a systematic and reproducible screening of the Materials Cloud 3D database (MC3D)~\cite{MC3D} for BCS superconductors.
Importantly, the MC3D has been obtained combining three crystal-structure databases (COD~\cite{Graulis2011}, ICSD~\cite{ICSD} and MPDS~\cite{MPDS}), filtering out as much as possible entries that are not backed by experimental results, and keeping only stoichiometric compounds at standard conditions. 
The goal here is to explore with state-of-the-art electronic-structure calculations known materials for novel properties, in the same spirit as Mounet et al.~\cite{Mounet2018}, rather than predicting novel materials (with all the challenges that it entails for novel properties~\cite{Cheetham2024}).
Starting from a set of 4533 non-magnetic metals (as predicted from Kohn-Sham DFT, that underestimates band gaps), we first perform an initial screening at progressively higher levels of precision using the \textsc{SSSP} pseudopotential library~\cite{Prandini2018} which also includes projector augmented wave method~\cite{Blochl1994} pseudopotentials. 
Based on these results, we select the top 949 materials for which we perform a new structural relaxation but this time using the 
\textsc{PseudoDojo} library~\cite{vanSetten2018} which contains exclusively norm-conserving pseudopotentials.
Out of these, we exclude 47 materials found to be magnetic after an additional test, 347 materials due to calculation failures or unconverged results, 270 showing unstable phonon modes, and 33 due to an insufficient quality in the interpolated band structures.
From the remaining 252 promising candidates, we compute Wannier functions, Eliashberg spectral functions $\alpha^2F(\omega)$ and the isotropic Migdal Eliashberg superconducting T$_{\rm c}^{\rm iso}$.
At that stage, 2 fail due to node failure, giving 250 T$_{\rm c}^{\rm iso}$. 
Finally, for the 144 materials with a T$_{\rm c}^{\rm iso}$ larger than 5~K, we perform a full anisotropic Migdal-Eliashberg superconducting calculation.
These final calculations are memory intensive and 4 fail for this reason, leaving a final set of 140 anisotropic T$_{\rm c}^{\rm aniso}$ predictions. 
For each material, our database, available openly on the Materials Cloud Archive~\cite{MCA2025} and presented in the Supplemental Material, reports electronic band structures and phonon dispersions, Eliashberg spectral functions $\alpha^2F(\omega)$, Allen-Dynes $T_{\rm c}$~\cite{Allen1975}, isotropic and anisotropic Migdal-Eliashberg superconducting gaps $\Delta_{n\mathbf{k}}$ and $T_{\rm c}$~\cite{Margine2013}.
To our knowledge, this effort represents the largest database of converged anisotropic superconducting Migdal-Eliashberg calculations.

From the list of 250 T$_{\rm c}^{\rm iso}$ candidates, we find that 82 of them are known superconductors with a $T_{\rm c}$ in
reasonable agreement with experiment when accounting for the experimental mismatch with the computed structure.
The remaining 168 compounds are not known to be superconducting, and 24 of these have a predicted $T_{\rm c}$ above 10~\si{\kelvin}.
To ensure the robustness of these predictions with respect to the (approximate) Kohn-Sham band structures and position of the Fermi energy (which in experiments can be affected by defects and unintentional doping or self-doping), we introduce a sensitivity criterion based on homogeneous doping within the rigid band approximation. 
We then discuss in detail three appealing examples: the hole-doped two-gap superconductor BaB$_2$, the potentially first half-Heusler superconductor ZrRuSb, and the perovskite TaRu$_3$C.

\section{Results} 
\subsection{Screening of the MC3D database and high-throughput calculations \label{sec:screening}}
\begin{figure}[t]
    \centering
    \includegraphics[width=0.9\linewidth]{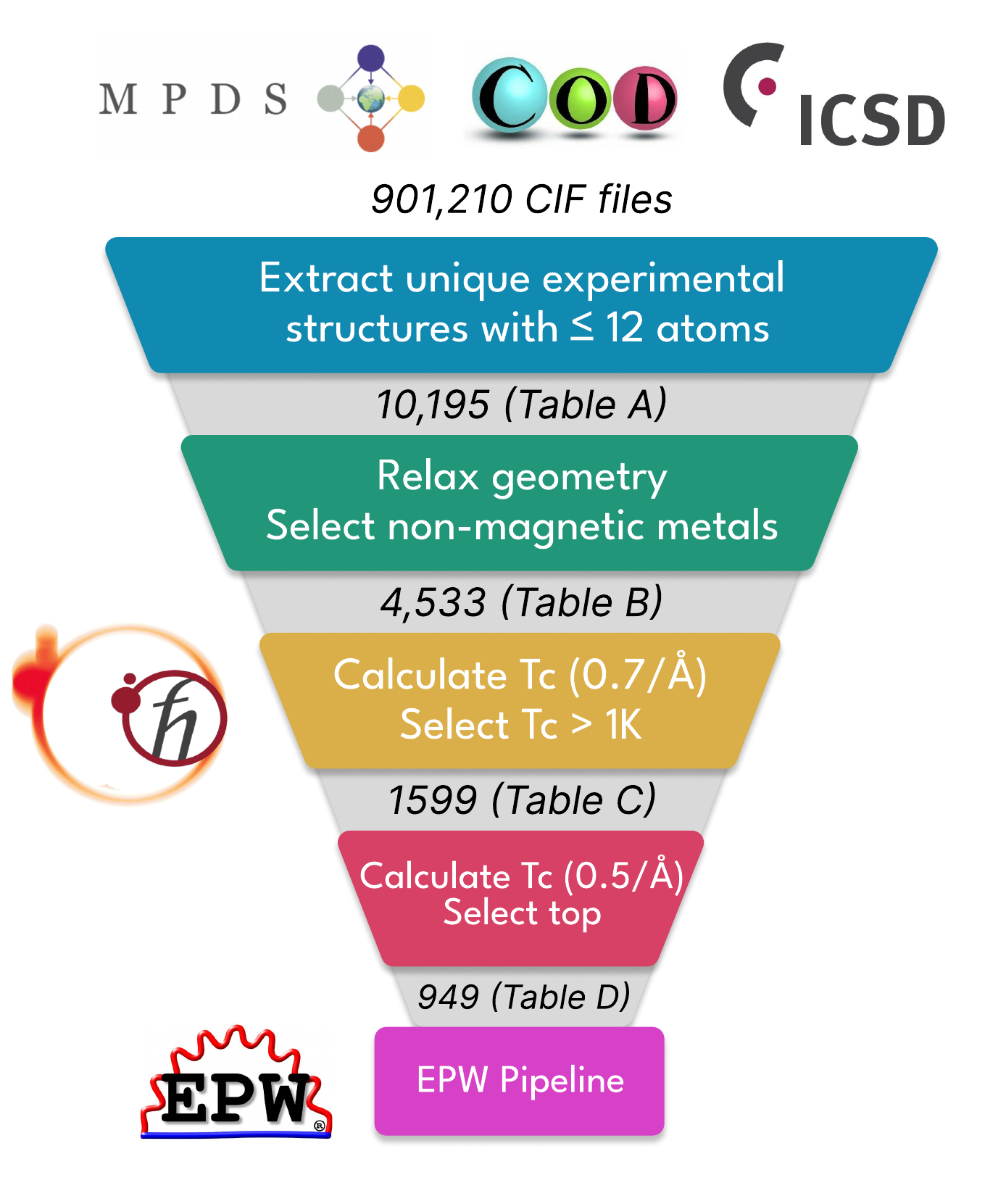}
    \caption{Initial screening of experimentally known materials.
    The structures are extracted from the \textsc{MPDS}~\cite{MPDS}, \textsc{COD}~\cite{Graulis2011}, and \textsc{ICSD}~\cite{ICSD} databases using the \textsc{Quantum ESPRESSO} electron-phonon coupling workflow.
    Table A, B, C, and D correspond to the full list of materials belonging to each category and published with this work~\cite{MCA2025}. }
    \label{fig:qe-screening}
\end{figure}

We start our screening from a set of Crystallographic Information File (CIF) files extracted from 3 databases available in the literature: \textsc{MPDS}~\cite{MPDS}, \textsc{ICSD}~\cite{ICSD}, and \textsc{COD}~\cite{Graulis2011}.
We have analyzed these extensively and combined them into a database of unique, experimentally known structures at ambient conditions.
Subsequently, each structure was optimized with DFT calculations following a strict protocol after initializing them in a high-spin ferromagnetic state to identify magnetic materials.
The optimized geometries are published as the Materials Cloud 3D database (\textsc{MC3D})~\cite{MC3D}, whose details are provided in Supplementary Sec.~1.
Starting from the \textsc{MC3D}, non-magnetic materials with 12 atoms or less in the primitive unit cell are considered (see Fig.~\ref{fig:qe-screening}).
Finally, metals are selected based on the Fermi level and occupations at the Kohn-Sham DFT level, leading to 4533 compounds that we screen for potential superconductivity.

Next, we calculate the electron-phonon interactions using density-functional perturbation theory as implemented in the \textsc{Quantum ESPRESSO} distribution~\cite{Giannozzi2017}.
In our initial screening, we created a workflow to calculate the Eliashberg spectral function and corresponding Allen-Dynes $T_{\rm c}^{\rm AD}$, progressively increasing the precision of the calculations by using denser and denser phonon ($\bq$) and commensurate electron ($\bk$) sampling grids.
In the first step, we use a coarse homogeneous $\bq$-point grid with density 0.7~\si{/\angstrom^{-1}} and identify structures with a $T_{\rm c}^{\rm AD}$ above 1~\si{\kelvin} as potential superconductors.
In the second iteration, a $\bq$-point grid with density 0.5~\si{/\angstrom^{-1}} is used from which we make a selection of the 949 structures with the highest $T_{\rm c}$, see the Methods section for additional details on the workflow and Supplementary Sec.~2 for details on the structures, soft modes, and failure rate.   
\begin{figure*}
    \centering
    \includegraphics[width=1\linewidth]{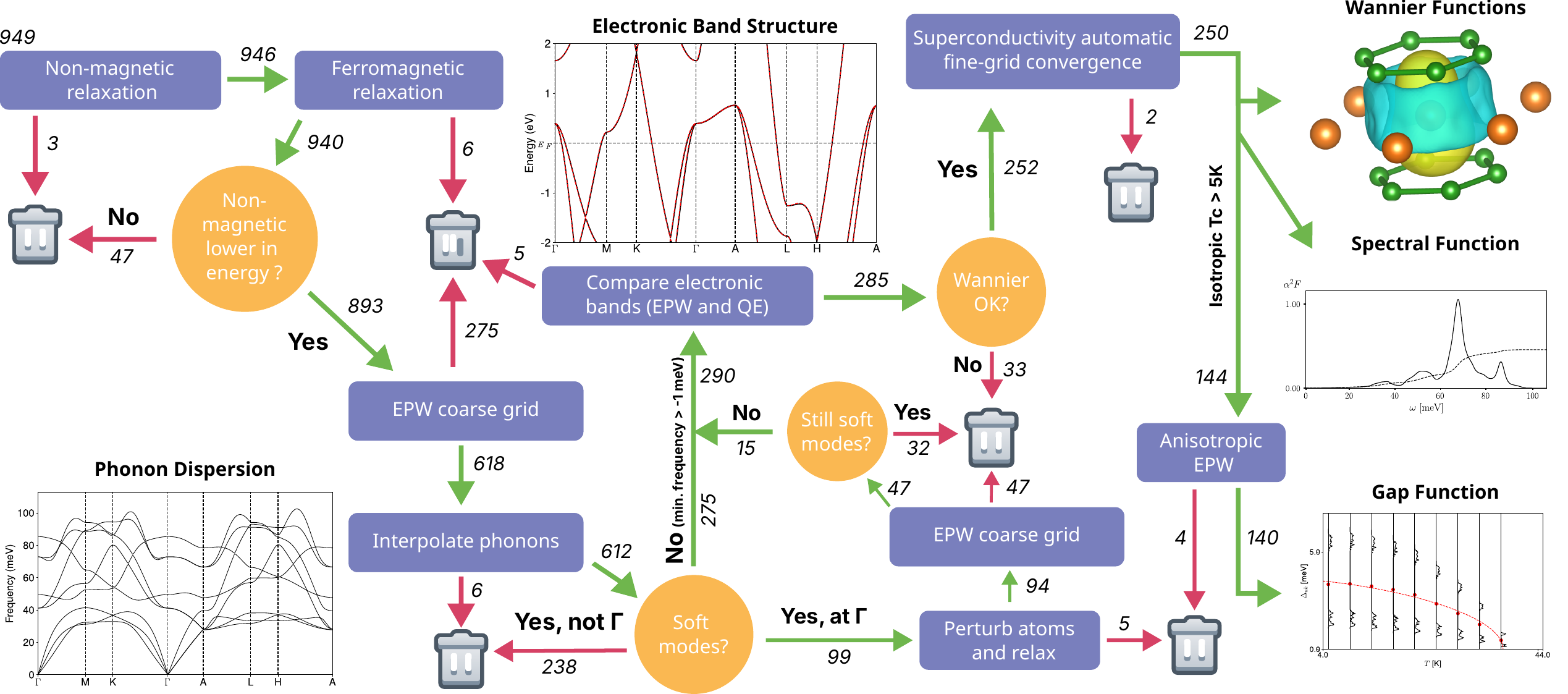}
    \caption{Flowchart of the Electron-Phonon Wannier (\textsc{EPW}) workflow.
    The pipeline starts from the top candidates obtained from the initial \textsc{Quantum ESPRESSO} screening. 
    The red arrows corresponds to structures that are discarded along the pipeline due to failed runs, magnetism, the presence of soft modes or insufficient quality of the Wannierization.}
    \label{fig:epw-pipeline}
\end{figure*}
Although this initial screening, based on the linear interpolation approach of Wierzbowska \textit{et al.}~\cite{Wierzbowska2005}, is a solid first step to detect materials as superconductors, a significant number of structures have a $T_{\rm c}^{\rm AD}$ that is not yet converged with respect to sampling, see Supplementary Sec.~2.
Since denser grids are computationally prohibitive and the Allen-Dynes approximation to the $T_{\rm c}$ has limited accuracy~\cite{Margine2013}, in the final step of our screening we use the \textsc{EPW} code~\cite{Ponce2016, Lee2023} to interpolate electron-phonon matrix elements on ultra-dense grids at low computational costs.
Thus, we take the top 949 candidates with the goal of first calculating the isotropic Eliashberg $T_{\rm c}$ using the \textsc{EPW} code (see Fig.~\ref{fig:epw-pipeline}).

Since magnetism aligns spins in a particular direction, it breaks the spin-singlet pairing of Cooper pairs~\cite{Eremin2020} and is detrimental to standard superconductivity. 
We therefore first perform both a non-magnetic and ferromagnetic structural relaxation of all the materials to determine if the ferromagnetic solution is more stable (in order to discard it), resulting in a set of 893 non-magnetic metals.    
To perform the interpolations with \textsc{EPW}, we rely on the SCDM-k method~\cite{Damle2018} to automatically generate maximally-localized Wannier functions that span the occupied and lowest unoccupied bands, and that act as an essentially exact interpolator of the Kohn-Sham band structures~\cite{Marzari2012}.
In the coarse grid calculations required to set up the Wannier interpolations, the workflows for 275 compounds failed, mostly due to issues related to reaching computational convergence or resource limits (out-of-memory errors or time limits), see Supplementary Section 2 for a detailed account of the error rates.  
For the remaining 618 coarse grid calculations, we interpolate the phonon band structure and apply an acoustic sum rule that enforces the 15 Born-Huang conditions~\cite{Lin2022} and then check the dynamical stability of the structure.

At this stage we exclude 6 materials due to non-Hermiticity of the dynamical matrices due to the dipole-dipole interatomic force constant ansatz~\cite{Gonze1997}; for the 612 interpolated phonon dispersions, we find that 238 have soft modes away from $\Gamma$ which indicates potential instabilities. 
We also find 99 structures with soft modes at $\Gamma$ that we try to stabilize by applying a random perturbation of the atomic positions and relaxing the geometries; this allows us to stabilize a further 15 structures. 
For the remaining 290 materials, we assess the quality of the Wannier functions interpolation by comparing an electronic band structure directly calculated with DFT and the interpolated one, accepting a maximum 50~\si{\milli\electronvolt} weighted band distance~\cite{Prandini2018}.
This leads us to 252 structures that are dynamically stable and have accurate Wannier interpolations of the electronic band structure.
These are then passed to the superconductivity workflow (see Fig.~\ref{fig:epw-pipeline}), which performs an automatic convergence of the interpolation mesh based on $T_{\rm c}^{\rm AD}$.
Once $T_{\rm c}^{\rm AD}$ is found to converge within 1~\si{\kelvin}, another run is performed to calculate the isotropic Eliashberg critical temperature $T_{\rm c}^{\rm iso}$ with the corresponding interpolation mesh.
During this automatic convergence of the interpolated momentum meshes, we exclude 2 materials due to convergence issues, which leaves 250 materials for which we perform a converged $T_{\rm c}^{\rm iso}$ calculation. 
For the 144 structures that have a $T_{\rm c}^{\rm iso}$ above 5~\si{\kelvin} we perform a final anisotropic Eliashberg $T_{\rm c}^{\rm aniso}$ calculation to obtain the superconducting gap $\Delta_{n\bk}(T)$ on the Fermi surface.
During the anisotropic Eliashberg calculations, 4 materials failed due to out-of-memory error or reaching execution time limits, leading to a final set of 140 high-quality, tightly converged anisotropic Eliashberg results.
Additional details on the \textsc{EPW} workflow are provided in the Methods section.  

\subsection{Analysis of the supercond-EPW database \label{sec:results}}

\begin{table}[h!]
  \caption{\label{table:known}
  Top 30 compounds that are known superconductors.
  From the screening the list is ordered according to the computed $T_c^{\rm aniso}$, together with chemical formulae, space group (SG) number, Allen-Dynes transition temperature ($T_c^{\rm AD,c}$ in K) computed with direct coarse grids, Allen-Dynes transition temperature ($T_c^{\rm AD,f}$ in K) interpolated on fine grids, isotropic Eliashberg transition temperature ($T_c^{\rm iso}$) in K, anisotropic Eliashberg transition temperature ($T_c^{\rm aniso}$) in K, and experimental $T{\rm _c}$.
The electron-phonon coupling strength $\lambda$ and logarithmic phonon frequency $\omega_{\rm log}$ entering the Allen-Dynes formula are reported in Supplementary Sec.~7.
All the predicted $T{\rm _c}$ are obtained with an effective Coulomb potential of $\mu^*=0.13$, a reasonable assumption for many experimentally verified superconductors.
All experimental references and full details of the structures and properties are provided in Supplementary Sec.~7. 
}
\begin{tabular}{ l@{\hspace{2pt}}c@{\hspace{2pt}}c@{\hspace{2pt}}c@{\hspace{2pt}}c@{\hspace{2pt}}c@{\hspace{2pt}}r}
  \toprule\\
Material & SG & $T_c^{\rm AD,c}$ & $T_c^{\rm AD,f}$ & $T_c^{\rm iso}$ & $T_c^{\rm aniso}$ & Experimental $T_c$   \\
\hline \\
 MgB$_2$  & 191 & 25.6 & 12.3 & 16.5 & 36.8 & 39.0~\cite{Nagamatsu2001}                     \\
 MoN      & 187 & 23.7 & 26.7 & 32.1 & 36.6 & 4.0-6.0~\cite{Ganin_2006}                     \\
 Nb$_3$Sn & 223 & 11.8 & 18.3 & 25.6 & 35.6 & 17.9-18.3~\cite{Stewart_2015,Xu_2017}         \\
 RuO$_2$  & 136 & 11.9 & 21.9 & 26.1 & 34.0 & $<$ 0.3~\cite{Lin_2004}                       \\
 PdH      & 225 & 30.0 & 21.2 & 25.9 & 30.1 & 8.5-11.0~\cite{Sko_kiewicz_1973,Harper_1974}  \\
 TaSe$_2$ & 194 &  4.1 &  8.9 & 20.1 & 27.5 & 0.1~\cite{Van_Maaren_1967}                    \\
 NbSe$_2$ & 194 &  5.8 & 16.3 & 21.9 & 26.7 & 5.7~\cite{Guillam_n_2008}                     \\
 V        & 229 & 19.0 & 20.4 & 27.1 & 26.1 & 4.7-5.4~\cite{Eisenstein_1954,Hamlin2015}     \\
 MoC      & 194 & 22.3 & 17.3 & 20.8 & 25.5 & 8.0~\cite{Toth_1968}                          \\
 V$_3$Pt  & 223 & 10.2 & 14.9 & 18.5 & 25.2 & 3.0~\cite{Ramakrishnan_1986}                  \\
 VC       & 225 & 18.9 & 17.2 & 20.7 & 25.0 & 1.8~\cite{Ziegler_1953}                       \\
 NbS$_2$  & 194 &  8.5 & 18.5 & 24.3 & 25.0 & 6.1~\cite{Witteveen_2021}                     \\
 TaS$_2$-2H  & 194 &  4.7 & 16.2 & 21.3 & 23.7 & 0.5-2.2~\cite{Murphy_1975,Navarro_Moratalla_2016} \\ 
 TaB$_2$  & 191 & 14.4 & 15.5 & 17.5 & 21.6 & $<$ 1.5~\cite{Rosner_2001}                    \\
 Ta$_3$Sn & 223 &  9.4 & 11.4 & 15.6 & 21.5 & 4.2-8.35~\cite{Courtney_1965,Wada_1973}       \\
 Nb       & 229 & 13.1 & 15.1 & 19.7 & 21.2 & 9.1-9.5~\cite{Cappelletti_1967,Matthias1963}  \\
 Mo$_3$Os & 223 & 16.3 & 15.4 & 18.9 & 20.8 & 7.3-12.7~\cite{Matthias_1955,Flukiger_1974}   \\
 CaC$_6$  & 166 & 10.5 & 11.6 & 12.1 & 19.1 & 11.5~\cite{Emery_2005}                        \\
 ZrN      & 225 & 10.3 &  9.8 & 12.6 & 18.6 & 9.3-9.6~\cite{Hardy_1954}                     \\
 ZrRuP    & 189 &  8.2 & 13.0 & 15.9 & 17.9 & 13.3~\cite{M_ller_1983}                       \\
 YCI      &  12 &  4.9 &  8.5 & 13.6 & 17.7 & 9.85~\cite{Simon_1996}                        \\
 ZnNi$_3$N& 221 &  8.4 & 14.5 & 18.3 & 17.6 & 3.0~\cite{Uehara_2010}                        \\
 Tc       & 194 & 11.9 & 12.7 & 16.3 & 17.1 & 7.9-11.2~\cite{Giorgi_1966,Matthias1963}      \\
 YCCl     & 12  &  3.9 &  8.0 & 12.0 & 14.9 & 2.3~\cite{Simon_1996}                         \\
 TiN      & 225 & 10.3 &  9.9 & 12.9 & 14.2 & 5.5-5.6~\cite{Hardy_1954,Hulm_1971}           \\
 NbB$_2$  & 191 &  9.0 & 10.6 & 13.0 & 13.8 & 0-5.0~\cite{Xiao_2006,Mudgel_2008}            \\
 HfN      & 225 & 10.4 &  9.7 & 11.7 & 13.7 & 5.8~\cite{Potjan_2023}                        \\
 BaGe$_3$ & 194 &  2.5 &  8.7 & 10.9 & 13.2 & 4.0-6.5~\cite{Fukuoka_2011,Castillo_2016}     \\
 ScSe     & 225 &  5.0 &  8.0 & 10.0 & 12.3 & 3.7~\cite{Moodenbaugh_1978}                   \\
 Sn       & 139 &  4.9 &  7.7 & 11.0 & 11.6 & 3.7~\cite{Matthias1963}                       \\
\botrule
\end{tabular}  
\end{table}

The resulting database of structures and properties, called \emph{supercond-EPW}, contains the primitive unit cells, electronic band structures, phonon dispersions, Eliashberg spectral functions $\alpha^2F(\omega)$, $T_{\rm c}^{\rm AD,f}$ on the fine grid, and $T_{\rm c}^{\rm iso}$ for 250 metals.
For 140 of these having a $T_{\rm c}^{\rm iso} > 5$~K, we also provide the $T_{\rm c}^{\rm aniso}$ and $\Delta_{n\bk}(T)$. 
These results are reported in Supplementary Sec.~7 and on the Materials Cloud Archive~\cite{MCA2025}.
To understand which of the structures among these candidates are known superconductors, we perform an extensive literature survey and search for reported critical temperatures.
We find that 137 out of 250 materials have been experimentally investigated for their critical temperature and, for 82 of them where the structure and not only the chemical formula are reported, we can confirm that the experimental structure matches the calculated one, see Supplementary Sec.~7.
From these, 8 are false positives (not superconducting down to 1~\si{\kelvin}).
%small mpds-S1323564, mpds-S1629062,  mpds-S1830331, mpds-S311773
%big  mpds-S1704431, mpds-S1815445, mpds-S451301, mpds-S541753
%
This means that 74 (90\%) of the structures for which we have a confirmed experimental result are true positives, which is a strong indication that our procedure can reliably identify superconducting materials from a large pool of initial candidates.
To assess the risk of removing materials erroneously (false negatives), we extracted a list of 32 well-known superconductors from the literature and investigated how many of these are discarded by our procedure (see Supplementary Sec.~3).
Of these, we missed seven; these include three failures in the phonon calculation (which could be resolved in future improvements of the workflows to make them more robust) and two due to an insufficient interpolation quality of the band structure (an issue that can be resolved via improved approaches to automated Wannierization using projectability disentanglement~\cite{Qiao2023}).
Finally, we find that the remaining two cases are due to the presence of soft modes in the phonon dispersions.
We highlight the particular case of $\delta$-NbN with an experimental $T_{\rm c}$ of 16~\si{\kelvin}~\cite{Shy1973} that was discarded due to soft modes at the $\mathbf{q}=\mathbf{X}$ point, the reason being that the experimental $\delta$-NbN can only be prepared with a small N deficiency and that it has been found that the disorder can be simulated with a large electronic smearing~\cite{Babu2019}.

In Table~\ref{table:known}, we list the top 30 compounds for which we have an experimental result and a matching structure, with the highest $T_{\rm c}^{\rm iso}$, along with the corresponding $T_{\rm c}^{\rm AD}$ temperatures for the coarse and fine grid, $T_{\rm c}^{\rm aniso}$ and reported experimental value.
In the list, several classes of well-known superconductors are present, such as elemental Nb and Tc, MgB$_2$, and the A15 phases (Nb$_3$Sn and Mo$_3$Os).
However, we also note some of the false positives mentioned above, RuO$_2$ and TaB$_2$, which have a high calculated $T_{\rm c}$, yet experimental reports indicate that they are not superconducting.
Moreover, in some cases we find a large discrepancy between the reported experimental critical temperature and the calculated ones.
These results indicate that comparing computational predictions with experimental data presents certain challenges. 
These include (i) the limited availability of comprehensive databases, (ii) uncertainties in the structures present in experiments,  and (iii) the presence of material complexities such as doping, grain boundaries, defects, strain, off-stoichiometries, or site-antisite mixing, all of which can influence $T_{\rm c}$ and may not be captured by state-of-the-art computational methods.
On the computational side, the most significant approximations are (i) the neglect of spin fluctuation~\cite{Kawamura2020} in our calculations, (ii) the use of a fixed Coulomb parameter $\mu^*=0.13$, and (iii) the use of Kohn-Sham band structure; DFT is not a spectral theory, and more expensive many-body perturbation theory should be employed to calculate electronic excitations~\cite{Reining2017}.
Removing experimental data with reported limitations and materials where spin fluctuation is known to dominate, 
we find that our calculated T$_c$ are in reasonable agreement with experimental data, see Sec.~3 of the Supplementary information for detailed report and comparison.

\begin{table}[t!]
  \caption{\label{tab:unknown}
Top 30 compounds from this screening that are predicted to be BCS superconductors.
Largest $T{\rm _c}$ (K) for materials predicted to be BCS superconductors, following the same convention as Table~1.
For $T_c^{\rm AD,f}$ and $T_c^{\rm iso}$ we provide a sensitivity analysis which gives the change of $T_{\rm c}$ upon homogeneous electron (superscript) and hole (subscript) doping of 10$^{21}$~cm${^{-3}}$.
Materials that have a potentially magnetic ground state based on self-consistent Hubbard calculations~\cite{Kulik2006}, or for which we could not confirm the structure or source, are noted.
BaB$_2$ is listed separately, as a hole doping of 0.065~holes per unit cell is required to stabilize the phase.
}
\begin{tabular}{ l@{\hspace{9pt}}r@{\hspace{9pt}}r@{\hspace{9pt}}r@{\hspace{9pt}}r@{\hspace{9pt}}r@{\hspace{9pt}}}
\toprule
Material & SG & $T_c^{\rm AD,c}$ & $T_c^{\rm AD,f}$ & $T_c^{\rm iso}$ & $T_c^{\rm aniso}$ \\
\hline 
W$_2$N$_3$   & 194 & 11.3 & 20.2$_{\scriptscriptstyle -0.9}^{\scriptscriptstyle +0.6}$ & 26.3$_{\scriptscriptstyle -1.0}^{\scriptscriptstyle +0.6}$  & 33.7   \\ [2pt]
Be$_2$B      & 225 &  9.7 &  7.7$_{\scriptscriptstyle -0.3}^{\scriptscriptstyle +0.5}$ & 10.9$_{\scriptscriptstyle +0.2}^{\scriptscriptstyle +0.7}$  & 22.1   \\ [2pt]  
PtO          & 131 &  8.9 & 13.3$_{\scriptscriptstyle -0.1}^{\scriptscriptstyle +0.1}$ & 16.3$_{\scriptscriptstyle -0.2}^{\scriptscriptstyle +0.3}$  & 19.0   \\ [2pt]
Mo$_3$Be     & 223 &  9.3 &  8.1$_{\scriptscriptstyle -0.4}^{\scriptscriptstyle +0.4}$ & 10.5$_{\scriptscriptstyle -0.4}^{\scriptscriptstyle +0.4}$  & 17.6   \\ [2pt]
TeAs         & 225 &  6.4 & 10.6$_{\scriptscriptstyle -0.4}^{\scriptscriptstyle +0.5}$ & 13.8$_{\scriptscriptstyle -0.6}^{\scriptscriptstyle +0.6}$  & 17.2   \\ [2pt]
TaCoSb       & 216 & 11.4 & 11.4$_{\scriptscriptstyle -2.7}^{\scriptscriptstyle -0.7}$ & 14.1$_{\scriptscriptstyle -3.6}^{\scriptscriptstyle -0.7}$  & 16.8   \\ [2pt]
TaS          & 187 &  8.4 &  9.3$_{\scriptscriptstyle +0.0}^{\scriptscriptstyle -0.8}$ & 11.3$_{\scriptscriptstyle -0.6}^{\scriptscriptstyle -1.0}$  & 13.6   \\ [2pt]
ZrRuSb       & 216 &  5.1 &  8.9$_{\scriptscriptstyle -0.7}^{\scriptscriptstyle +0.5}$ &  9.9$_{\scriptscriptstyle -0.9}^{\scriptscriptstyle +0.5}$  & 13.2   \\ [2pt]
YIr$_3$B$_2$ &  12 &  6.2 &  4.8$_{\scriptscriptstyle +0.2}^{\scriptscriptstyle -0.2}$ &  8.6$_{\scriptscriptstyle +0.3}^{\scriptscriptstyle -0.3}$  & 12.2   \\ [2pt]
Zr$_2$Al     & 140 &  3.0 &  5.7$_{\scriptscriptstyle -0.2}^{\scriptscriptstyle +0.2}$ &  7.4$_{\scriptscriptstyle -0.0}^{\scriptscriptstyle +0.4}$  &  7.8   \\ [2pt]
MoB          &  63 &  3.9 &  4.8$_{\scriptscriptstyle -0.1}^{\scriptscriptstyle -0.1}$ &  6.8$_{\scriptscriptstyle -0.1}^{\scriptscriptstyle -0.1}$  &  7.7   \\ [2pt]
HfRuSb       & 216 &  3.8 &  4.9$_{\scriptscriptstyle -0.5}^{\scriptscriptstyle +0.5}$ &  6.6$_{\scriptscriptstyle -0.6}^{\scriptscriptstyle +0.6}$  &  6.4   \\ [2pt]
             & \multicolumn{4}{c}{Potentially magnetic} \\ [2pt]
\hline 
TaRu$_3$C    & 221 & 15.0 & 16.5$_{\scriptscriptstyle +1.4}^{\scriptscriptstyle -1.9}$ & 21.0$_{\scriptscriptstyle +4.4}^{\scriptscriptstyle -2.6}$ & 25.0   \\ [2pt]
NbRu$_3$C    & 221 & 15.6 & 17.5$_{\scriptscriptstyle +1.2}^{\scriptscriptstyle -1.5}$ & 21.8$_{\scriptscriptstyle +1.7}^{\scriptscriptstyle -2.0}$ & 24.9   \\ [2pt]
NbCoSb       & 216 & 12.4 & 16.3$_{\scriptscriptstyle +1.0}^{\scriptscriptstyle -1.5}$ & 20.4$_{\scriptscriptstyle +1.4}^{\scriptscriptstyle -1.7}$ & 21.4   \\ [2pt]
IrS$_2$      & 205 & 12.6 & 13.2$_{\scriptscriptstyle -0.0}^{\scriptscriptstyle -0.3}$ & 16.1$_{\scriptscriptstyle -0.0}^{\scriptscriptstyle -0.3}$ & 20.8   \\ [2pt] 
ZrS          & 129 & 10.9 & 10.7$_{\scriptscriptstyle -0.1}^{\scriptscriptstyle +0.0}$ & 13.4$_{\scriptscriptstyle -0.2}^{\scriptscriptstyle -0.3}$ & 18.8   \\ [2pt] 
Ti$_3$SnH    & 221 &  7.9 & 11.7$_{\scriptscriptstyle -0.3}^{\scriptscriptstyle -1.4}$ & 14.6$_{\scriptscriptstyle -1.4}^{\scriptscriptstyle -2.8}$ & 18.7   \\ [2pt]
RhS$_2$      & 205 & 16.6 &  9.4$_{\scriptscriptstyle +1.0}^{\scriptscriptstyle -0.6}$ & 11.8$_{\scriptscriptstyle +0.7}^{\scriptscriptstyle +0.5}$ & 17.3   \\ [2pt]
CrH          & 225 &  9.8 & 10.6$_{\scriptscriptstyle +0.3}^{\scriptscriptstyle -0.2}$ & 13.4$_{\scriptscriptstyle +0.3}^{\scriptscriptstyle -0.2}$ & 16.9   \\ [2pt]
TiRuSb       & 216 &  7.5 & 10.1$_{\scriptscriptstyle -1.0}^{\scriptscriptstyle +0.7}$ & 12.9$_{\scriptscriptstyle -1.9}^{\scriptscriptstyle +1.1}$ & 13.6   \\ [2pt]
RhSe         & 194 &  7.7 &  8.3$_{\scriptscriptstyle +1.0}^{\scriptscriptstyle -1.1}$ & 10.8$_{\scriptscriptstyle +0.8}^{\scriptscriptstyle -1.6}$ & 11.9   \\ [2pt] 
Ti$_2$Ga     & 194 &  3.9 &  5.2$_{\scriptscriptstyle -0.2}^{\scriptscriptstyle +0.1}$ &  7.1$_{\scriptscriptstyle -0.4}^{\scriptscriptstyle -0.1}$ &  8.3   \\ [2pt]
& \multicolumn{4}{c}{Unconfirmed source} \\ [2pt]
\hline 
YC           & 225 & 15.4 & 13.4$_{\scriptscriptstyle -0.9}^{\scriptscriptstyle +0.7}$ & 17.1$_{\scriptscriptstyle -0.9}^{\scriptscriptstyle +1.0}$  & 22.0  \\ [2pt]
TaMo$_2$B$_2$& 127 &  9.5 &  9.7$_{\scriptscriptstyle +0.6}^{\scriptscriptstyle -0.7}$ & 12.1$_{\scriptscriptstyle +0.5}^{\scriptscriptstyle -0.8}$  & 14.3  \\ [2pt]
Zr$_3$Sn     & 223 &  8.8 &  7.7$_{\scriptscriptstyle +0.3}^{\scriptscriptstyle -0.3}$ &  9.9$_{\scriptscriptstyle +0.2}^{\scriptscriptstyle -0.3}$  & 14.2  \\ [2pt]
LiAl$_2$Ge   & 225 &  3.8 &  8.0$_{\scriptscriptstyle -0.0}^{\scriptscriptstyle +0.1}$ &  9.7$_{\scriptscriptstyle -0.1}^{\scriptscriptstyle +0.1}$  & 10.8  \\ [2pt]
Li$_2$AlGe   & 216 &  6.2 &  5.8$_{\scriptscriptstyle -0.2}^{\scriptscriptstyle +0.3}$ &  7.3$_{\scriptscriptstyle -0.2}^{\scriptscriptstyle +0.3}$  &  9.3  \\ [2pt] 
AgF$_2$      &  14 &  8.1 &  5.1$_{\scriptscriptstyle +1.0}^{\scriptscriptstyle -0.7}$ &  6.7$_{\scriptscriptstyle +0.5}^{\scriptscriptstyle -1.4}$  &  7.5  \\ [2pt] 
& \multicolumn{4}{c}{Hole-doped stabilization} \\ [2pt]
\hline 
BaB$_2$      & 191 & 20.4 & 21.4$_{\scriptscriptstyle -0.1}^{\scriptscriptstyle +0.2}$ & 33.0$_{\scriptscriptstyle -0.2}^{\scriptscriptstyle +0.3}$  & 61.6   \\[2pt]
\botrule
\end{tabular}
\end{table}

In Table~\ref{tab:unknown}, we report the top 30 experimentally known compounds that are predicted
by the workflows discussed here to be BCS superconductors.
We note that VC, CrH, and Be$_2$B have also been reported as superconductors in the computation work of Choudhary et al.~\cite{Choudhary2022}, with predicted $T_{\rm c}$ of 28.1~K, 10.7~K, and 8.8~K, respectively. 
These values are in good agreement with the 20.7~K, 13.4~K, and 10.9~K isotropic $T_{\rm c}$ that we compute.
In addition, W$_2$N$_3$ monolayer~\cite{Campi_2021} and TaMo$_2$B$_2$~\cite{Chen2024} have also been proposed theoretically with predicted $T_{\rm c}$ of 21~K and 12~K, respectively. 
The other materials presented in Table~\ref{tab:unknown} are novel in the sense that they have never been reported as superconductors.
However, many materials in the list contain heavy elements with partially occupied $d$ or $f$ shells and could be magnetic, even if the initial screening found them to be non-magnetic.
Henceforth, we additionally validate all transition-metal compounds in Table~\ref{tab:unknown} employing state-of-the-art self-consistent relaxations using on-site and inter-site Hubbard correction~\cite{Timrov2021} for the materials in their nonmagnetic and ferromagnetic configurations, using the \textsc{HP} code~\cite{Timrov2022} to determine $U$ and $V$ from first-principles. 
The Hubbard-corrected band structure of the materials and their magnetic energy landscape is reported in Supplementary Sec.~4 and the materials with a ferromagnetic ground state are noted in Table~\ref{tab:unknown}.
Since magnetism is detrimental to BCS superconductivity~\cite{Eremin2020}, all results are reported for materials in their PBE nonmagnetic ground state.

Overall, the material with the highest predicted $T_{\rm c}^{\rm aniso}$ is BaB$_2$, which displays interesting two-gap superconductivity, reminiscent of the well-known behavior observed in isostructural and isoelectronic MgB$_2$~\cite{Nagamatsu2001}.
One could wonder how such a simple binary material had not been discovered already. 
As discussed in the next section, we find that pristine BaB$_2$ displays a small instability and relaxes to a different structure than the experimentally reported one.
However, upon p-doping of 0.065 holes per unit cell, the MgB$_2$-like phase stabilizes and shows high promise.
Despite these results, careful consideration is necessary since DFT electronic band structure and Fermi level position are inexact~\cite{Marzari2021}.
We therefore introduce a sensitivity analysis for the predictions, which consists in computing the change of Allen-Dynes $T{\rm _c}$ and isotropic $T{\rm _c}$ upon a homogeneous electron ($+$) and hole ($-$) doping of 10$^{21}$~cm${^{-3}}$.
This doping corresponds to a medium/high metallic doping that can realistically be achieved through substitution or interstitial doping (through ion implantation, the Fermi level can be experimentally fine-tuned at \si{\milli\electronvolt} level to achieve the optimal doping to enhance the superconducting critical temperature~\cite{DePalma2024}).
Doping levels above 10$^{22}$~cm${^{-3}}$ are usually not feasible due to potential phase segregation~\cite{Grovogui2021}. 
We translate the doping density into energy shifts around the Fermi level, where changes of $T_{\rm _c}$ can be evaluated via the Eliashberg equations (see Supplementary Sec.~5 for a detailed description). 
We report in Table~\ref{tab:unknown} this sensitivity analysis and find 13 superconductors for which $T{\rm _c}$ changes by less than 0.5~K upon $\pm 10^{21}$~cm${^{-3}}$ doping, and thus can be considered more robust with respect to the predictions.

\begin{figure*}[t]
    \centering
    \includegraphics[width=0.93\linewidth]{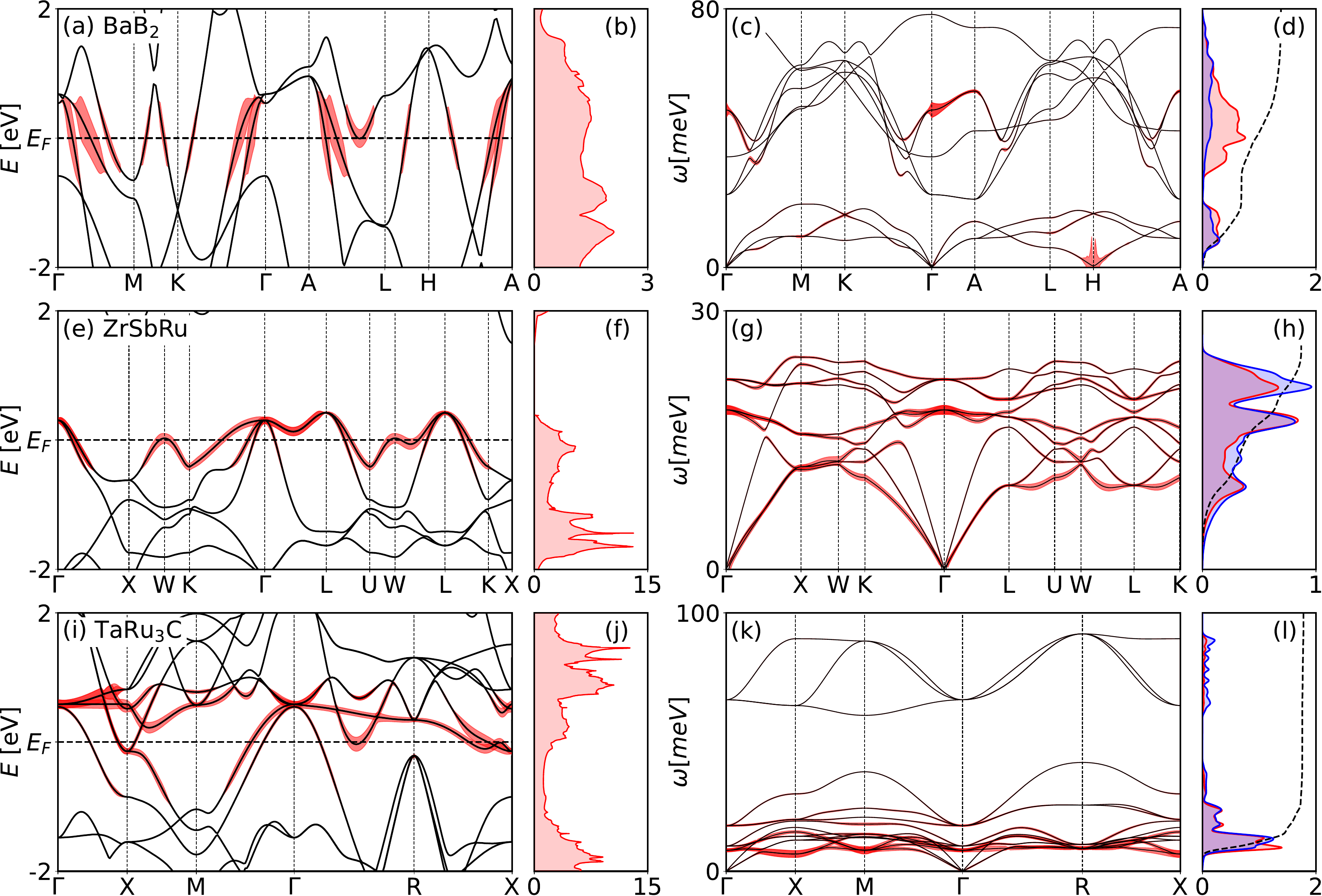}
    \caption{
    Analysis of electronic, vibrational, and electron-phonon properties of BaB$_2$, ZrSbRu, and TaRu$_3$C.
    Electron band structures, electron density of states, phonon dispersions, spectral functions~(red), accumulated spectral functions~(black dashed) and phonon density of states~(blue) for doped-BaB$_2$~(a-d), ZrRuSb~(e-h) and TaRu$_3$C~(i-l). 
    The band-resolved electron-phonon coupling constant $\lambda_{n\textbf{k}}$ and mode-resolved electron-phonon coupling constant $\lambda_{\textbf{q}\nu}$ are plotted on the band structures and phonon dispersion as proportional to the linewidth. 
    The largest band-resolved coupling constants are 2.82, 1.11, 3.16, respectively.}
    \label{fig:epa}
\end{figure*}

An interesting class of structures in Table~\ref{tab:unknown} are the half-Heusler compounds NbCoSb, TaCoSb, TiRuSb, ZrRuSb, and HfRuSb~\cite{Evers1997}.
Many properties of Heusler compounds~\cite{Winterlik2008}, including the electronic structure, are typically related to their valence electron count (VEC).
Half-Heusler is one of the most common structure types for ternary intermetallics, and commonly half-Heusler phases have a VEC of 8, 18, or 28 per formula unit, corresponding to a closed-shell configuration. 
Therefore, half-Heusler phases are typically semiconducting, whereas the few open-shell half-Heusler phases are typically metallic in nature~\cite{Anand2018}.
Although many examples of full-Heusler superconductors have been discussed in the literature, both in experimental~\cite{Graf2011} and computational~\cite{Hoffmann2023} work, superconductivity in half-Heuslers is found to be non-conventional~\cite{Timm2017, Xiao2018} or topological in nature~\cite{Tafti2013} and related to their non-centrosymmetric structure~\cite{Smidman2017}.
For full-Heusler phases, Graf \textit{et al.}~\cite{Graf2011} have proposed a relationship between the VEC and superconducting properties, citing a prevalence of superconducting full Heuslers with VEC 27.
For TiRuSb, HfRuSb, ZrRuSb, NbCoSb, and TaCoSb the VEC is 17 and 19, indicating an electron deficiency and surplus compared to VEC 18, respectively.
Reports on conventional superconductivity in half-Heusler compounds are scarce, making this an exciting avenue for further exploration.
There are also several perovskite structures in the list, including TaRu$_3$C, ZnNi$_3$N, NbRu$_3$C, Ti$_3$TlN and Ti$_3$SnH, where TaRu$_3$C has the highest predicted  $T_c^{\rm aniso}$ of 25.0~\si{\kelvin}.
Perovskites~\cite{Kim2022} and antiperovskites~\cite{Hoffmann2022} are well-known families of superconductors where empirical relations exist between $T{\rm _c}$ and cation charges in the perovskites cages~\cite{Poole2000}.

\subsection{Analysis of  BaB$_2$, ZrSbRu, and TaRu$_3$C}\label{sec:discussion}

\begin{figure*}[t]
    \centering
    \includegraphics[width=0.93\linewidth]{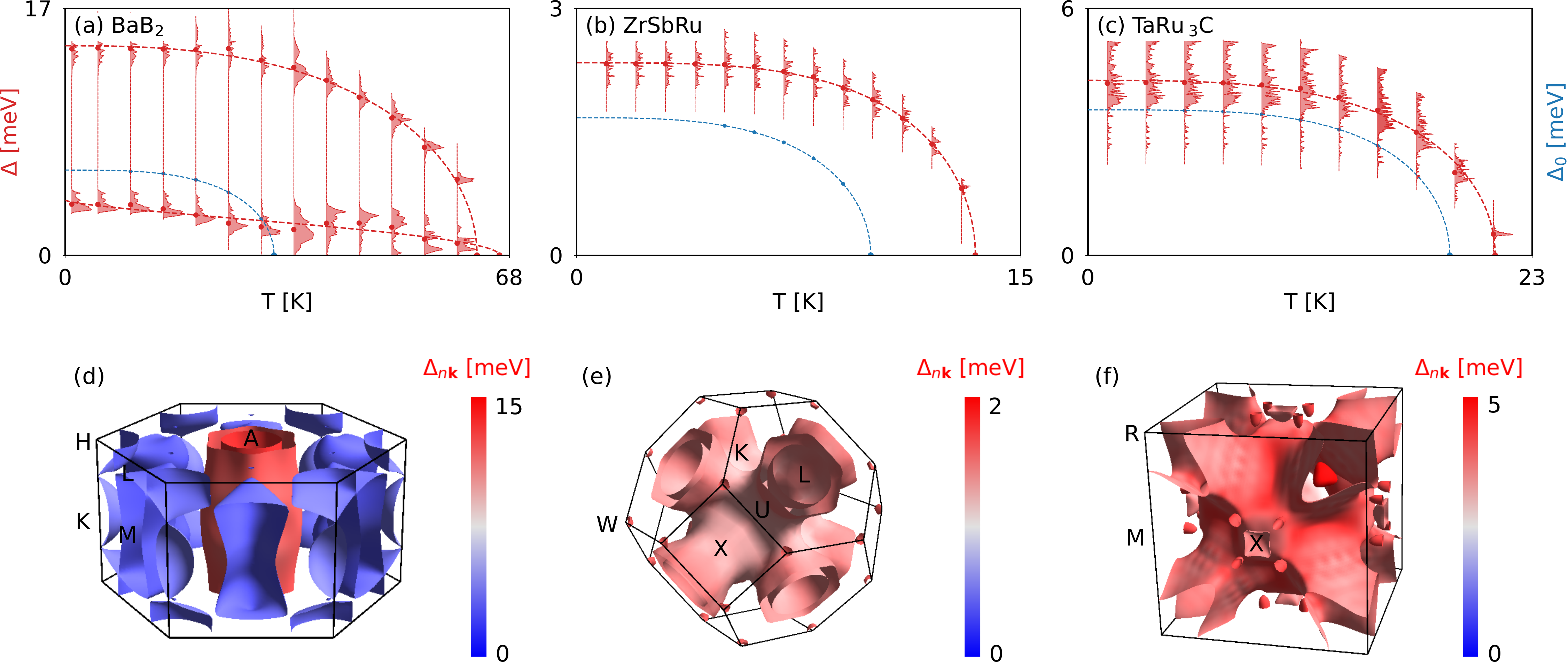}
    \caption{
    Analysis of superconducting properties of BaB$_2$, ZrSbRu, and TaRu$_3$C. 
    Isotropic gap functions $\Delta_0$ (blue), histograms of anisotropic gap functions $\rho(\Delta, T)$ (red), superconducting gaps on Fermi surfaces for doped-BaB$_2$~(10~\si{\kelvin})~(a,d), ZrRuSb~(5~\si{\kelvin})~(b,e) and TaRu$_3$C~(5~\si{\kelvin})~(c,f).}
    \label{fig:sc}  
\end{figure*}

In this section, we focus specifically on BaB$_2$, the ZrRuSb half-Heusler, and TaRu$_3$C, as these constitute interesting candidates.
For these three materials, we have performed more detailed calculations using the intermediate representation (IR)~\cite{Mori2024} of the Migdal Eliashberg implementation in the \textsc{EPW} software.
Figures~\ref{fig:epa} and~\ref{fig:sc} show their electronic and vibrational properties, spectral function, isotropic and anisotropic superconducting gap function $\Delta_{0}(i\pi T)$ and $\Delta_{n\mathbf{k}}(i\pi T)$.
For the anisotropic gap function, we show the histogram of the superconducting gap function $\rho(\Delta)$ at the lowest Matsubara frequency $\Delta_{n\mathbf{k}}(i\pi T)$, defined as~\cite{Mori2024}
\begin{equation}
    \rho(\Delta(T))=\sum_{n\mathbf{k}} \delta(\Delta_{n\mathbf{k}}(i\pi T)-\Delta(T))\delta(\epsilon_{n\mathbf{k}}-\epsilon_{\rm F}),
\end{equation}
where $\epsilon_{\rm F}$ is the Fermi level.
The temperature-dependent isotropic gap $\Delta_0(T)$ and anisotropic gap $\Delta_{n\mathbf{k}}(i\pi T)$ are fitted by the BCS-type gap function $\Delta^{\rm BCS}(T)=\Delta^{\rm BCS}(0) \sqrt{1-(T/T{\rm _c})^\beta}$. 
To further analyze the structure of the electron-phonon coupling, we compute the mode-resolved and band-resolved coupling constants $\lambda_{\mathbf{q}\nu}$ and $\lambda_{n\mathbf{k}}$.

We start with a closer investigation of BaB$_2$, which is an isostructural and isoelectronic counterpart of the well-known two-gap superconductor MgB$_2$ that was experimentally discovered more than two decades ago~\cite{Nagamatsu2001}.
Although BaB$_2$ has been investigated in the context of superconductivity~\cite{EmetereMoses2020}, its critical temperature has never been determined nor predicted. 
Based on the results from our high-precision workflow described in Fig.~\ref{fig:epw-pipeline}, the pristine hexagonal phase is unstable due to a soft mode at the $H$ point. 
Upon further investigation, we find that homogeneous hole doping above 0.065/u.c. (1.46$\cdot$10$^{21}$~cm~$^{-3}$) can stabilize the pristine phase while superconductivity remains unaffected up to at least 0.1/u.c. (see Supplementary Sec.~6).
Analogously to MgB$_2$, BaB$_2$ is a strongly coupled material with $\lambda=1.39$. 
From Fig.~\ref{fig:epa}(a-d), one can see that the electron-phonon coupling is highly anisotropic, with a dominant optical phonon mode at $\Gamma$ and an acoustic mode at $H$, which is unstable in the undoped structure. 
The interpolated isotropic transition temperature is $T_c^{\rm iso}$=32.0~\si{\kelvin}, larger than the $T_c^{\rm AD}$=24.4~\si{\kelvin}.
According to Fig.~\ref{fig:sc}(a,d), the Fermi surfaces are composed of two nested cylinders along $\Gamma$-$A$ line, with a superconducting gap of 11.54~\si{\milli\electronvolt} and pockets centered at $H$, $M$ and $K$ points with a superconducting gap of 1.50~\si{\milli\electronvolt}.
This corresponds to two anisotropic critical temperatures $T_c^{\rm aniso, 1}$=61.6~\si{\kelvin} and $T_c^{\rm aniso, 2}$=60.6~\si{\kelvin}, larger than the isotropic approximation due to the strong anisotropy. 
In Fig.~\ref{fig:epa}(a), the band-resolved coupling strength also shows that the coupling is dominated by the two nested cylindrical Fermi-surface sheets.

We continue with the five half-Heusler phases NbCoSb, TaCoSb, HfRuSb, TiRuSb, and ZrRuSb studying their magnetic states in more detail. 
The magnetic landscape is probed using a random forest search where magnetic configurations are constrained and then relaxed using electronic structure calculations at the PBE and PBE+U level with the \textsc{ROMEO} code~\cite{Ponet2024}.
The magnetic search is performed in the conventional supercell (12 atoms) to allow for more complex magnetic states. 
The algorithm is stopped when the ratio of newly discovered unique states to the number of trials goes below 0.2. 
We show the magnetic energy landscape for each configuration for the four half-Heuslers in Supplementary Sec.~4 and find that all are non-magnetic at the PBE level and ZrRuSb is the only one that remains non-magnetic also at the PBE+U level.
We therefore investigate ZrRuSb further and find that its Fermi surface is composed of a sphere centered at $\Gamma$~(invisible in Fig.~\ref{fig:sc}(e)), two nested connected neck-like formations at the L points, and small spheres centered at the W points. 
According to Fig.~\ref{fig:epa}(e,g,h), the electron-phonon coupling is overall isotropic, dominated by the phonon density of states, through enhancement of the optical $\Gamma$ phonon and acoustic phonons near X and L.
The interpolated isotropic and anisotropic transition temperatures are 9.9~\si{\kelvin} and 11.1~\si{\kelvin}, respectively.
However, we note in the DOS of Fig.~\ref{fig:epa}(f) that the high predicted superconductivity in ZrRuSb comes from the position of the Fermi level being located near a van Hove singularity with a flat band around the $W$ $\mathbf{k}$-point.
This means that this prediction might not be robust upon small changes of the Fermi level resulting from DFT inaccuracies or experimental conditions. 
This is supported by the sensitivity check shown in Table.~\ref{tab:unknown}, where ZrRuSb varies by over 1~K upon medium doping level.
Finally, we study the perovskite TaRu$_3$C as our highest-$T_c$ (potentially magnetic) material from Table~\ref{tab:unknown}.
Indeed, all these materials are non-magnetic at the DFT level but magnetic at the PBE+U level, which tends to over-stabilize magnetic solutions.
For TaRu$_3$C, we show in Fig.~\ref{fig:epa}(k,l) that only low-energy phonons contribute to the electron-phonon couplings, due to the large phonon density of states at these energies. 
TaRu$_3$C also possesses very strong electron-phonon couplings, manifested by $\lambda=1.77$. 
The Fermi surface in Fig.~\ref{fig:sc}(f) is composed by sphere at $\Gamma$ (invisible in the figure) and connected neck-like formations at $R$ and really small pockets near $X$. 
From Fig.~\ref{fig:epa}(i), we note that the Fermi surfaces at the $\Gamma$-$R$ line and $R$-$X$ lines, although they coupled strongly with the phonons, are really sensitive to the shift of Fermi energy.
The $T_c^{\rm iso}$=21.0~\si{\kelvin} and $T_c^{\rm aniso}$=25.0~\si{\kelvin} indicate the somewhat isotropic nature of this material, which is also shown by the homogeneous Cooper pair density on the Fermi surface in Fig.~\ref{fig:sc}(f).
We have shown that these three materials discussed here (and many others identified in this study) host intriguing new physics that is worth further experimental investigation.

In conclusion, starting from 4533 experimentally known compounds with 12 atoms or less in the primitive unit cell obtained from the Materials Cloud 3D database and determined to be non-magnetic metals at the DFT (PBEsol) Kohn-Sham level, we have performed a high-throughput search for novel superconductors using a combination of the \textsc{Quantum ESPRESSO}, Wannier90, and \textsc{EPW} codes.
Of the top 250 candidates, 137 have been investigated for superconductivity; for the 82 where we can positively match structures, 74 are also reported as superconductors, giving a 90\% ratio of true positives and demonstrating the ability of our approach to flag materials as potential superconductors.
Of the remaining 113 that have so far not been reported in the experimental literature, we have 24 with a predicted $T_{\rm c}^{\rm aniso}$ above 10~\si{\kelvin}.
Among the novel candidates, there is the double gap superconductor BaB$_2$ with a predicted $T_{\rm c}$ of 61.6~\si{\kelvin}, the half-Heusler ZrRuSb with a 11.1~K critical temperature, higher than the 12 known full-Heusler superconductors present in our dataset,  and the perovskite TaRu$_3$C with a predicted $T_{\rm c}$ of 25.0~\si{\kelvin}.
If experimentally confirmed, ZrRuSb would be the first half-Heusler superconductors with a VEC of 17 and 19 to exhibit conventional superconductivity.
The present results demonstrate the potential of high-throughput calculations to identify new superconductors and provide a valuable resource for future experimental studies, highlighting not only the predictive power of high-throughput calculations but also their limits.
Computationally, this work sets the stage for exploring these databases further, including also larger unit cells or higher accuracy methods Wannierization~\cite{Qiao2023} or for electronic band structures, potentially unveiling overlooked superconductors with intriguing physical properties and promising application potential.

\section{Methods}\label{sec:Methods}

\subsection{Initial interpolation workflow}\label{sec:workflows-qe-elph}

Our initial electron-phonon workchain is a workflow that calculates the Eliashberg spectral function based on the linear interpolation approach of Wierzbowska et al.~\cite{Wierzbowska2005}.
It consists of the following five steps: (i) a DFT calculation performed on a fine \textbf{k}-point that is used later to perform the linear interpolation, (ii) a second DFT calculation on a coarser grid required to calculate the phonons in the next step, (iii) a phonon calculation which calculates the electron-phonon coefficients on the coarse \textbf{k}-grid with a commensurate \textbf{q}-grid, (iv) the calculations of the real-space force constants, and (v) the Fourier interpolation over a dense \textbf{q}-grid and to calculate the final electron-phonon coupling and corresponding spectral function on the fine \textbf{k}-point grid.

We use pseudopotentials from the standard solid-state pseudopotentials (SSSP) PBEsol efficiency v1.1~\cite{Prandini2018} library.
For each material, we take the highest value for the plane-wave cutoffs among the suggested values for each of the elements present.
We use a \textbf{k}-point grid that is twice the density of the \textbf{q}-grid.

\subsection{EPW interpolation workflow}

\begin{figure}[ht]
    \centering
    \includegraphics[width=0.99\linewidth]{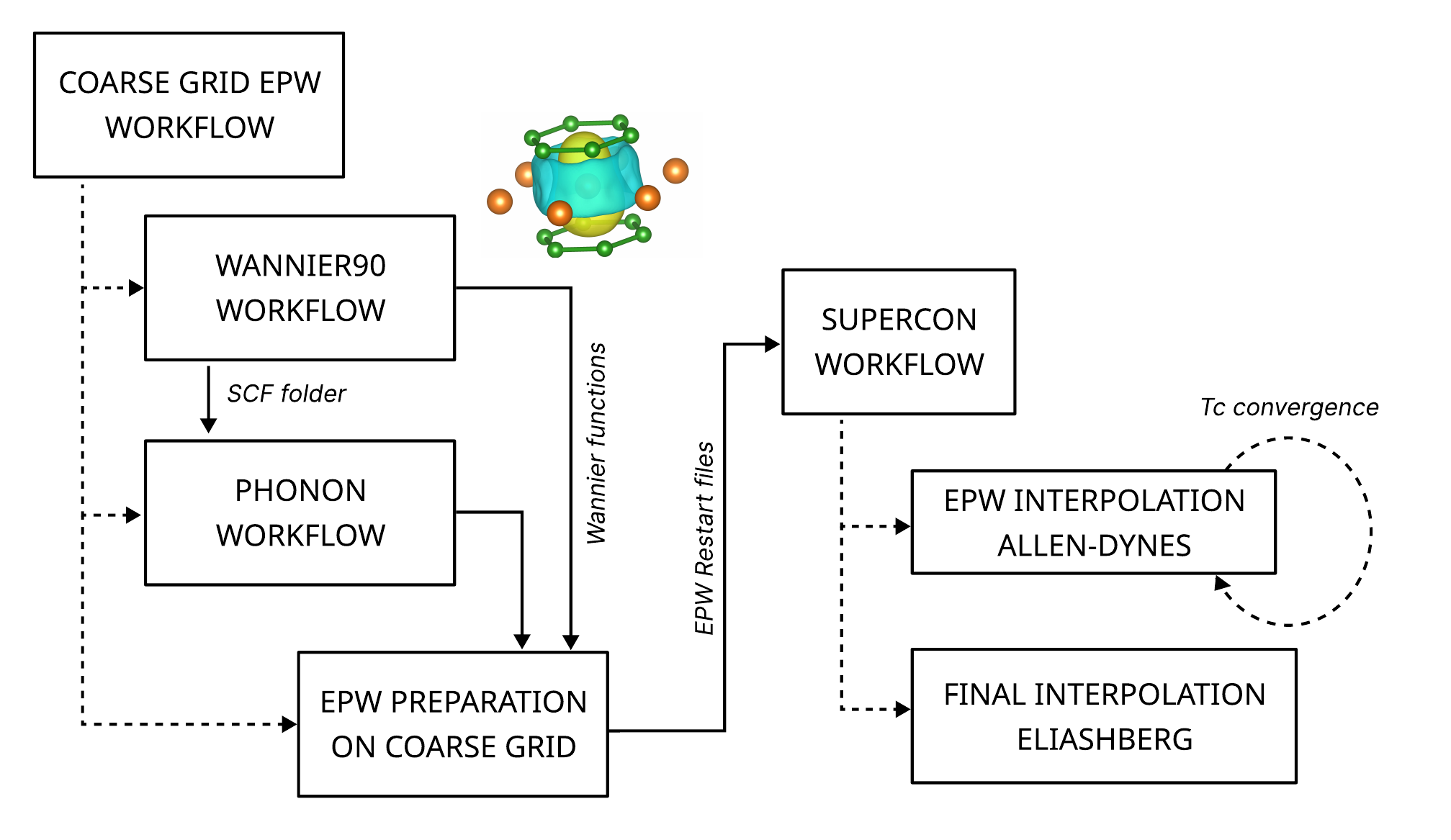}
    \caption{Schematic of the superconducting workflow. 
    Schematic of the steps executed to obtain the isotropic Eliashberg $T_c$ using the \textsc{EPW} workflows implemented in \textsc{AiiDA}.}
    \label{fig:workflow-epw}
\end{figure}

The \textsc{EPW} pipeline described in the main text uses several components that run \textsc{Quantum ESPRESSO} in combination with \textsc{EPW} to calculate the electron-phonon coupling and the superconducting critical temperature at a high precision, see Fig.~\ref{fig:workflow-epw} for a schematic representation of the workflow.

The \textsc{EPW} workchain takes care of the coarse-grid \textsc{Quantum ESPRESSO}~\cite{Giannozzi2017}, \textsc{Wannier90}~\cite{Pizzi2020} and \textsc{EPW}~\cite{Ponce2016,Lee2023} calculations that are required for the subsequent interpolations.
The workflow consists of the following steps: (i) a first workflow that constructs the Wannier functions for the input structure using the SCDM-k method~\cite{Damle2018}, (ii) a phonon calculation is performed to compute the dynamical matrices and perturbed potentials, and (iii)
 the input files required for the \textsc{EPW} calculation are converted into the expected format and an \textsc{EPW} calculation is run on the coarse grid to produce the restart files for the interpolation, which are stashed on the remote machine in a permanent location.
For the EPW pipeline, we switch to the norm-conserving PBE pseudopotentials from the scalar-relativistic table of \textsc{PseudoDojo} v0.5~\cite{vanSetten2018} as projector augmented wave method pseudopotentials have not been extensively tested with the \textsc{EPW} code.
For the energy cutoffs, we select the highest hint value for each element and once again take the highest one from the set of atoms contained in each material. 

Based on the stashed restart files produced by the first workchain, a second superconducting workchain is run to interpolate the electron-phonon coupling on dense momentum grids and calculate the superconducting critical temperature.
The main feature of this workflow is that it automatically converges the fine-grid interpolation meshes for both the \textbf{k} and \textbf{q} points, based on a calculation of the Allen-Dynes critical temperature and a specified threshold (default 1~\si{\kelvin}).
Once at least 3 calculations have been performed in the convergence loop and the Allen-Dynes critical temperature has converged, the workflow runs a final calculation on the fine grid using the full Eliashberg theory to calculate the isotropic Eliashberg critical temperature on the converged fine grid.

\subsection{Data availability and author contributions}\label{sec:Data}

All data generated or analysed during this study are included in this Article, its Supplementary Information, and the Materials Cloud Archive~\cite{MCA2025}.

\paragraph*{Contributions:}
S.P. and N.M. conceived the project, 
M.B. and S.P. designed the AiiDA workflows,
M.B. performed high-throughput calculations with AiiDA,
Y.Z. performed analysis of stability, sensitivity and detailed superconductivity,
G.T., A.G.G. and M.B performed literature search,
L.B. performed magnetic calculation with PBE+U+V and the ROMEO code,
J.Q. helped with the automatic Wannier function generation,
F.v.R. provided experimental guidance,
S.P., G.P., E.C., and N.M. supervised the work. 
All authors contributed to the writing of the manuscript.
These authors contributed equally: Marnik Bercx, Samuel Poncé.

\begin{acknowledgments}
We acknowledge financial support from the NCCR MARVEL, a National Centre of Competence in Research, funded by the Swiss National Science Foundation (grant number 205602), as well as the European Centre of Excellence MaX “Materials design at the Exascale” (grant no. 824143).
The work is also supported by a pilot access grant from the Swiss National Supercomputing Centre (CSCS) on the Swiss share of the LUMI system under project ID “PILOT MC EPFL-NM 01”, a CHRONOS grant from the CSCS on the Swiss share of the LUMI system under project ID “REGULAR MC EPFL-NM 02”, and a grant from the CSCS under project ID mr32.
S. P. and Y. Z. acknowledge support from the Fonds de la Recherche Scientifique de Belgique (FRS-FNRS) and T.W011.23 (PDR-WEAVE) and also supported by the Walloon Region in the strategic axe FRFS-WEL-T.
M.B. and G.P. gratefully acknowledge support from the SwissTwins project, funded by the Swiss State Secretariat for Education, Research and Innovation (SERI).
G.P. gratefully acknowledges support from the Swiss National Science Foundation (SNSF) Project Funding (grant 200021E\_206190 ``FISH4DIET''), F.v.R acknowledges support from SNSF Eccellenza Funding (grant PCEFP2\_194183).
L.B. and N.M. gratefully acknowledge support from the Deutsche Forschungsgemeinschaft (DFG) under Germany’s Excellence Strategy (EXC 2077, No. 390741603, University Allowance, University of Bremen) and Lucio Colombi Ciacchi, the host of the “U Bremen Excellence Chair Program”.
Computational resources have been provided by the PRACE award granting access to Discoverer in SofiaTech, Bulgaria (OptoSpin project id. 2020225411), and by the EuroHPC JU award granting access to MareNostrum5 at Barcelona Supercomputing Center (BSC), Spain (Project ID: EHPC-EXT-2023E02-050), and by the Consortium des Équipements de Calcul Intensif (CÉCI), funded by the FRS-FNRS under Grant No. 2.5020.11 and by the Walloon Region, as well as computational resources awarded on the Belgian share of the EuroHPC LUMI supercomputer.
\end{acknowledgments}

\end{document}

\title[Article Title]{Supplemental Information: Charting the landscape of Bardeen--Cooper--Schrieffer superconductors in experimentally known compounds}

\author[1,*]{Marnik Bercx}
\author[2,3,**]{Samuel Ponc\'e}
\author[2]{Yiming Zhang}
\author[4]{Giovanni Trezza}
\author[5]{Amir Ghorbani Ghezeljehmeidan}
\author[6]{Lorenzo Bastonero}
\author[7]{Junfeng Qiao}
\author[8]{Fabian O. von Rohr}
\author[1]{Giovanni Pizzi}
\author[4]{Eliodoro Chiavazzo}
\author[1,6,7]{Nicola Marzari}

\affil[1]{\scriptsize Laboratory for Materials Simulations (LMS), Paul Scherrer Institut (PSI), CH-5232 Villigen PSI, Switzerland} % \affpsi

\affil[2]{\scriptsize European Theoretical Spectroscopy Facility, Institute of Condensed Matter and Nanosciences, Université catholique de Louvain, Chemin des Étoiles 8, B-1348 Louvain-la-Neuve, Belgium} 
\affil[3]{\scriptsize WEL Research Institute, avenue Pasteur, 6, 1300 Wavre, Belgium} 
\affil[4]{\scriptsize Department of Energy, Politecnico di Torino, Italy} 
\affil[5]{\scriptsize Electronic Components, Technology and Materials (ECTM), TU Delft, The Netherlands}
\affil[6]{\scriptsize Bremen Center for Computational Materials Science, and MAPEX Center for Materials and Processes, University of Bremen, D-28359 Bremen, Germany}  % \affbremen
\affil[7]{\scriptsize Theory and Simulation of Materials (THEOS) and National Centre for Computational Design and Discovery of Novel Materials (MARVEL), \'Ecole Polytechnique F\'ed\'erale de Lausanne (EPFL), CH-1015 Lausanne, Switzerland} % \afftheos
\affil[8]{\scriptsize Department of Quantum Matter Physics, University of Geneva, 24 Quai Ernest-Ansermet, CH-1211 Geneva, Switzerland} % \affgeneva

\affil[*]{marnik.bercx@psi.ch}
\affil[**]{samuel.ponce@uclouvain.be}

\maketitle

\section{MC3D database of inorganic materials}\label{sec:si-mc3d}

The MC3D~\cite{MC3D} is a database of structures optimized with the \textsc{Quantum ESPRESSO} code~\cite{Giannozzi2009, Giannozzi2017} using fully-automated workflows developed in \textsc{AiiDA}~\cite{Uhrin2021, Huber2020}.
The starting set of structures for the geometry optimization is obtained from the COD~\cite{Graulis2011}, the ICSD~\cite{ICSD} and the MPDS~\cite{MPDS} databases.
Each CIF file is parsed via an \textsc{AiiDA} workflow that removes unnecessary tags, performs minor corrections to the syntax, and parses the contents to extract the corresponding structure.
Subsequently, the parsed structures are normalized and transformed in their primitive unit cell using the \textsc{SeeK-path}~\cite{Hinuma2017} library, and a uniqueness analysis is performed to remove duplicate structures.
Finally, hydrogen-containing structures from the COD are removed as a result of the prevalence of molecular crystals in this database.
This pruning results in 79\,854 crystal structures. 
From these, we select all structures with a maximum of 12 atoms in the primitive unit cell that, according to the source database, are experimentally observed at ambient conditions, which yield 9\,929 structures.
For these, we perform crystal relaxation using \textsc{Quantum ESPRESSO}~\cite{Giannozzi2009, Giannozzi2017} and the PBEsol exchange-correlation functional~\cite{Perdew2008}.
Based on the occupations and magnetic moments obtained from the geometry optimization, we then select the non-magnetic metals to be screened for potential superconductors, resulting in a set of 4533 materials. 

Although various checks and fixes are performed on the extracted CIF files and parsing into actual structures, the structures are not verified with the original source papers during extraction, as doing so automatically is currently not possible.
Although we trust databases to extract structures from the literature accurately, we acknowledge that these processes, often semiautomated, may introduce occasional errors. 
Hence, while collecting the data for the top tables we present in the main text, we extracted the original papers from the source database and verified that the reported structures correspond to the ones we ran for our automated setup.
Problematic cases were discovered and are reported with an explanation in Sec.~\ref{sec:si-epw-results} below.

The analysis of Sec.~\ref{sec:si-epw-results} reveals several challenges and opportunities when using material databases such as the MPDS and ICSD as sources for high-throughput computational studies.
A significant issue is the availability of source papers, with four entries indicating missing or inaccessible references, often due to broken links or outdated metadata.
This hinders the ability to verify crystal structures, which is a critical step for reliable computational results.
Verified structures are explicitly noted for 23 entries, but mismatches and ambiguities, such as inconsistent formulas or undefined structures like solid solutions, add further complexity.

In the case of BaB$_2$, the only report is from Ref.~\cite{Torkar1972}.
After extensive additional simulations and given the great enthusiasm after the discovery of superconductivity in MgB$_2$, we suspect that pure BaB$_2$ cannot be made easily.
However, we find that BaB$_2$ could be stabilized via hole doping, see Sec.~\ref{sec:BaB2} below.  
YC has been reported to form in the rocksalt (SG 225) structure, however it is found to be heavily carbon-deficient (YC$_{1-x}$ with $x \approx$ 0.55) because full stoichiometry leads to structural instability, and the system compensates for charge imbalance by forming carbon vacancies, which help stabilize the cubic phase~\cite{atoji1969crystal}.
In the case of the predicted superconductivity in Zr$_3$Sn, this zirconium-tin alloy has indeed been fround to crystalize in an A15 structure, however it has been found to form as Zr$_4$Sn, with some Zr in the Sn position~\cite{kwon1992influence,samanta2021experimental}. 
For RhSe, the paper deposited on the MPDS is incorrect but we were able to find the correct one, see Sec.~\ref{sec:si-epw-results}. 
For LiAl$_2$Ge, we cannot find any source for this material. It is reported as a 225 Heulser phase but does not possess typical Heusler characteristics.
In the case of Li$_2$AlGe, the reported synthesis yield instead LiAlGe which is consistent with the assumptions of the doctoral thesis by Simone Zürcher (2001, ETH Zurich), namely that Li$_2$AlGe does not exist, but that it is LiAlGe with some intercalated Li~\cite{zurcher2001systematic}.
Finally, for AgF$_2$ it is reported in the MPDS entry S1219456 as space group 14. 
We find reports of AgF$_2$ with space group 60, but not with space group 14~\cite{muller2005structure}.
These issues highlight the difficulty of relying on databases for well-defined input structures without additional validation.
Despite these limitations, databases like MPDS and ICSD provide immense value by centralizing decades of experimental data, even when full details are not available.
Maintaining such resources is not a trivial task, requiring ongoing updates and validation.
Although challenges remain, these databases are indispensable tools for materials research and collaboration between users and developers will continue to enhance their utility and accuracy.

\section{Convergence with momentum grids and workflow error rate}

\begin{figure}[ht]
\centering
\includegraphics[width=0.8\linewidth]{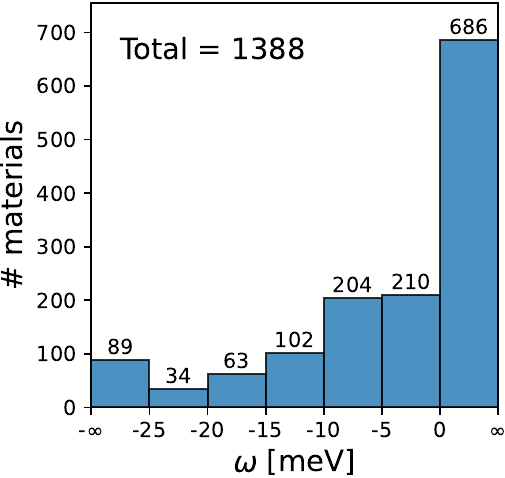}
\caption{
Lowest frequency phonon mode for the 0.5/\AA\, dataset containing 1599 materials from Fig. 1 of the main manuscript.
These calculations have been performed and relaxed with the SSSP PBEsol efficiency v1.1 pseudopotential library~\cite{Prandini2018}.
From these 1599 materials, 211 did not complete, mostly due to the calculations of phonon not converging. 
Out of the 1388 successful calculations, 492 materials have soft modes smaller than -5~meV. 
}
\label{fig:sssp}
\end{figure}

\begin{figure}[ht]
    \centering
    \includegraphics[width=0.9\linewidth]{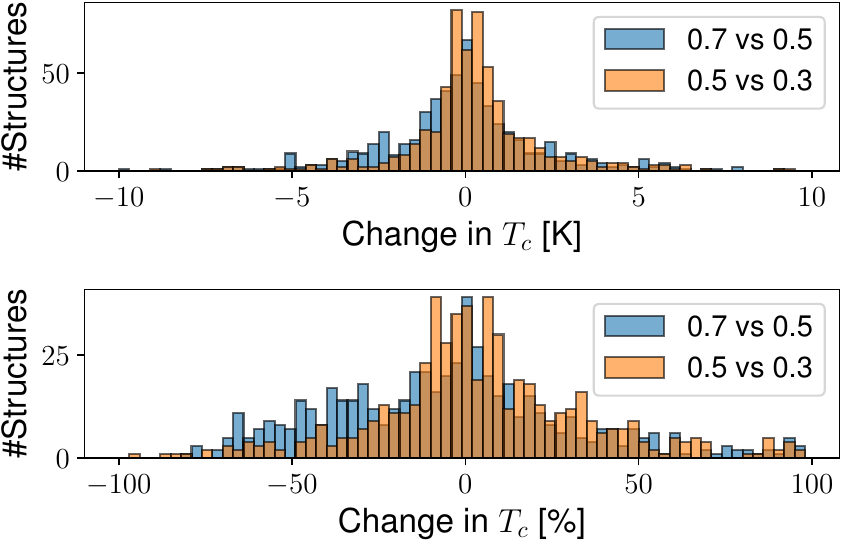}
    \caption{
    Convergence of $T_{\rm c}$ with momentum grid density.
    Change in absolute value and percentage of the Allen-Dynes $T_c$ with respect to the \textbf{q}-points density used in the calculation using the double-grid linear interpolation method as implemented in \textsc{Quantum ESPRESSO}.}
    \label{fig:convergence-qe}
\end{figure}

\begin{figure}[ht]
\centering
\includegraphics[width=0.8\linewidth]{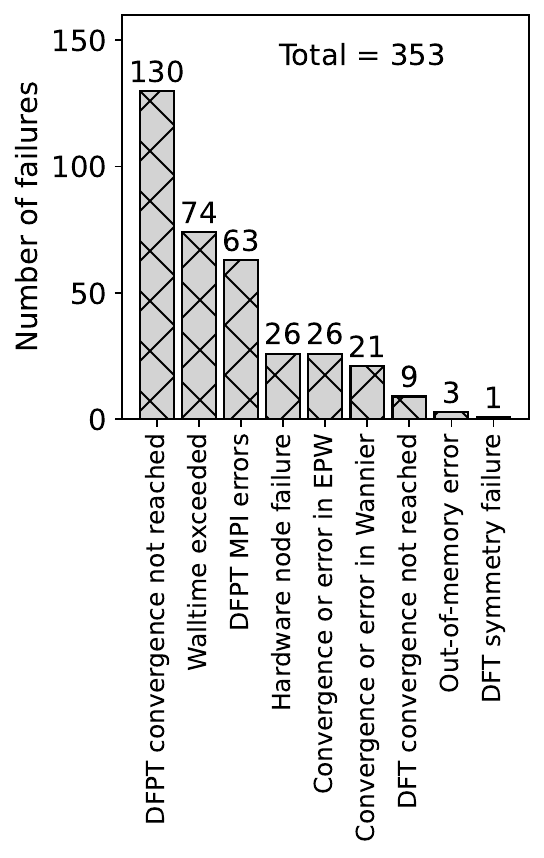}
\caption{\label{fig:errortype}
Analysis of types of failure for the \textsc{EPW} workflow shown in Fig.~2 of the main manuscript.
The errors in Fig.~2 of the manuscript are indicated with a red number in parenthesis. 
Among the 960 materials studied, we have a total of 379 errors. 
}
\end{figure}

Using the initial screening procedure described in the method section, we use double-grid interpolation to calculate the electron-phonon coupling and the Allen-Dynes critical temperature.
From the 0.5/\AA\, dataset containing 1599 materials,
we have 211 materials for which the DFPT calculations did not finish, mostly due to the calculations of the phonon not converging.
As can be seen in Fig.~\ref{fig:sssp}, some of the remaining 1388 materials have soft modes. 
Since soft modes can disappear with more converged results, we did not filter on soft modes and
took the 645 materials with the highest T$_c$ and computed denser 0.3/\AA\, grids (not discussed in the main manuscript).
In that case, the 56 structures failed due to convergence or node failure, giving us a total of 589 phonons for the 0.3/\AA\, set.

In Fig.~\ref{fig:convergence-qe}, we show the convergence of the Allen-Dynes $T_c$ with respect to the density of the \textbf{q}-points mesh used in the calculation.
We can see that even when using a quite computationally expensive \textbf{q}-point mesh of 0.3~\si{\angstrom}$^{-1}$, the Allen-Dynes $T_c$ is still not fully converged.
In fact, the change in the absolute value of Allen-Dynes $T_c$ is still of a similar value when compared to the difference between 0.5~\si{\angstrom}$^{-1}$ and 0.7~\si{\angstrom}$^{-1}$.
Using denser meshes would be prohibitively expensive for high-throughput screening, which motivates our switch to using the \textsc{EPW} code~\cite{Ponce2016, Lee2023} to obtain improved precision at a lower computational cost.

Finally, considering the dataset of 949 structures used for \textsc{EPW} calculations from Fig.~2 of the main manuscript, we find that the most common error is that the DFPT calculations using the \textsc{ph.x} code do not converge during the solution of the Sternheimer linear equation, see Fig.~\ref{fig:errortype}. 
These often signal a more fundamental hurdle including unstable structures, unstable ground-state calculations, and/or algorithmic instabilities. 
Instead, the second-largest error category are calculations where the walltime was exceeded. 
These could potentially be solved by running for longer or with more cores. 
The third largest category is related to message passing interface (MPI) errors, which relate to communications between cores during a calculation. 
Some of these errors are due to one node failing or to some deeper issues with the calculations.

\section{Comparison with experiment}

In order to verify the effectiveness of our computational approach, we search for results in the literature related to superconductivity for each material and classify them in one of the following categories:
\begin{itemize}
    \item \texttt{unknown}: No other results are found connecting this material to superconductivity.
    \item \texttt{known-mention}: The material is mentioned in the context of superconductivity, but no $T_{\rm c}$ is reported.
    \item \texttt{known-comp}: No experimental result is found, but it has been predicted theoretically to be a superconductor.
    \item \texttt{known-exp-formula}: We find an experimental result in the literature for the same composition, but we cannot verify that it is the same structure.
    \item \texttt{known-exp-structure}: We find an experimental result in the literature and can with reasonable confidence state that it is for the same structure as the one we calculated.
    \item \texttt{problematic}: There is some issues related with this materials.
    The issue is specified in the "Note" given in Sec.~\ref{sec:si-epw-results}.
\end{itemize}
Note that for structures classified as \texttt{known-exp}, the result from the literature can also be negative, e.g. ``not found to be superconducting down to 1.5~\si{\kelvin}''.
The table~1 in the main text only reports materials in the \texttt{known-exp-structure} category, as this provides the most reliable comparison for known superconductors.
Table~2 in the manuscript reports all computationally discovered materials which means any material that is not discarded nor has a corresponding experimental result in the literature (i.e. \texttt{unknown}, \texttt{known-comp} and \texttt{known-mention}).

To verify our true/false positive/negative rate, we restrict our results to materials classified as \texttt{known-exp-structure} and make a comparison between the computational and experimental result to determine the reliability of our predictions.
We flag a material as a "theoretically" superconducting in case the isotropic Eliashberg $T_{\rm c}$ is above 1~\si{\kelvin}.
For the experimental results, we consider a material to be superconducting in case any critical temperature is reported.
Conversely, a material is considered experimentally non-superconducting if no critical temperature is found down to a temperature below 1~\si{\kelvin}.
Technically, the structure \textit{could} still be superconducting at lower temperatures, but we consider this to be a reasonable cutoff for our analysis.
Based on the 82 structures that are classified as \texttt{known-exp-structure}, 74 of these are true positives, and 8 are found to be false positives.
That gives our approach a 90\% success rate in identifying superconductors starting from a list of structures.

\begin{table}[ht]
\centering
\caption{\label{table:known-super-wiki}
\textbf{List of commonly known stoichiometric BCS superconductors.}
List of known superconductors taken from Ref.~\cite{superconductors_wikipedia}, comparing the reported $T_c$ with the calculated isotropic Eliashberg values from our pipeline, or the reason the final result is not present.
}
\begin{tabular}{llrr}
\textbf{Name} & \textbf{Source} & \textbf{Exp. $T_c$[K]}  & \textbf{Calc. $T_c$ [K]} \\
\hline
In      & I-639816   & 3.4 & 6.1 \\
MgB$_2$ & M-S1637289 & 39 & 16.5 \\
Nb      & M-S1300517 & 9.26 & 19.7 \\
Nb$_3$Sn& M-S378780  & 18.3 & 25.6 \\
NbO     & M-S1800042 & 1.38 & 3.8 \\
Pb & M-S451186 & 7.19 & 4.2 \\
Re & M-S261398 & 2.4 & 7.7 \\
Sn & C-9012005 & 3.72 & 11.0 \\
Tc & M-S529740 & 7.46-11.2 & 16.3 \\
Ti & M-S1020386 & 0.39 & 6.1 \\
TiN & M-S313390 & 5.6 & 12.9 \\
Tl & M-S1020386 & 2.39 & 6.1 \\
V & I-60614 & 5.03 & 27.1 \\
YB$_6$ & M-S531588 & 8.4 & 8.8 \\
Zr & M-S553904 & 0.55 & 5.3 \\
ZrN & M-S1828487 & 10 & 12.6 \\
\hline
Cd & M-S1627329 & 0.52 & QE-0.7 $<$ 1 K \\
Ga & M-S1936861 & 1.083 & QE-0.7 $<$ 1 K \\
Hf & M-S534919 & 0.165 & QE-0.7 $<$ 1 K \\
Ir & M-S1500149 & 0.14 & QE-0.7 $<$ 1 K \\
Zn & M-S1825578 & 0.855 & QE-0.7 $<$ 1 K \\
$\alpha$-W & M-S1221477 & 0.015 & QE-0.7 $<$ 1 K \\
Os & M-S1933843 & 0.65 & QE-0.5 = 1.34 K \\
Ru & M-S260283 & 0.49 & QE-0.5 = 1.16 K \\
Mo & M-S452158 & 0.92 & QE-0.5 = 1.13 K \\
\hline
Al & M-S1930161 & 1.20 & node failure \\
Nb$_3$Al & M-S1714821 & 18 & max iterations  \\
Nb$_3$Ge & M-S555906 & 23.2 & max iterations  \\
$\alpha$-Hg & M-S1251834 & 4.15 & Wannier \\
Ta & M-S1928790 & 4.48 & Wannier \\
NbN & M-S313389 & 16 & Soft modes \\
$\beta$-Hg & M-S1006790 & 3.95 & Soft modes \\
\hline
\end{tabular}
\end{table}

In the list of structures classified as \texttt{known-exp-structure}, there is only one true negative and 3 false negatives.
This is a rather limited number of structures from which to draw any conclusions regarding false negatives. 
Moreover, this analysis would be biased for discussing true or false negatives as the results are only based on a pre-selected list of materials obtained from our preliminary screening with the \textsc{Quantum ESPRESSO} software.
Hence, we construct a list of well-known superconductors taken from Ref.~\cite{superconductors_wikipedia}.
After only selecting for BCS superconductors and removing doped or partially occupied materials, we obtain a list of 32 structures that were among our initial list of non-magnetic metals.
In Table~\ref{table:known-super-wiki}, we list the structures, their experimental $T_c$, and the calculated one or the reason why it was eliminated in our screening.
The most common reason for a false negative is the fact that the structure was eliminated during the \textsc{Quantum ESPRESSO} screening, either because the Allen-Dynes $T_c$ at the 0.7 precision level was below 1~\si{\kelvin} (6 cases) or it was not among the top candidates at the 0.7 precision level (3 cases).
Considering all these structures have an experimental $T_c$ below 1~\si{\kelvin}, or close to it, this is not necessarily a failure of our approach, 
Three structures were eliminated because of an issue in the \texttt{ph.x} calculation, two of them since they reached the maximum number of iterations, one because of a node failure.
Two structures, \ce{NbN} and $\beta$-\ce{Hg}, have non-$\Gamma$ soft modes in the phonon dispersion.
Finally, two more structures had an insufficient quality of the Wannierization based on our metric.

\begin{figure}[H]
\centering
\includegraphics[width=0.8\linewidth]{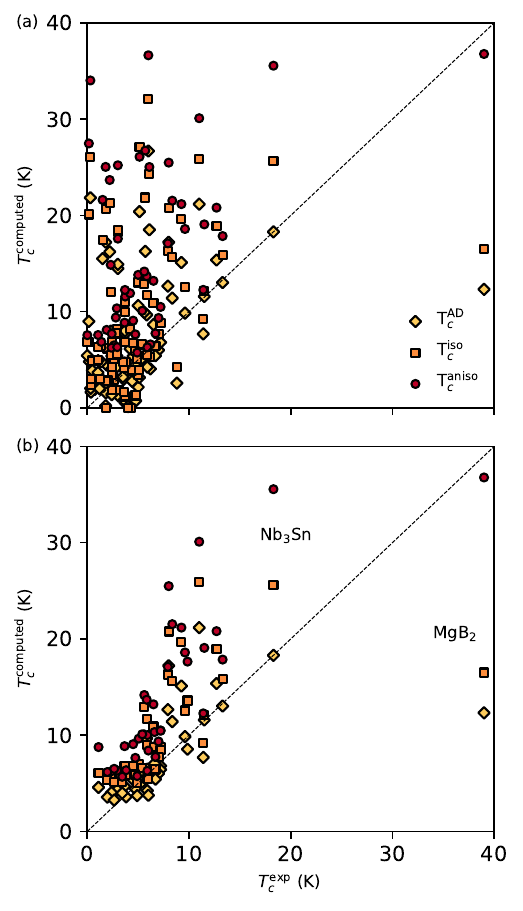}
\caption{
Comparison between computed and experimental superconducting critical temperature T$_c$.
(a) All 82 materials where experimental value were found, and (b) curated set of 37 materials where we removed materials with potential theoretical or experimental limitations. 
}
\label{fig:comparison_exp}
\end{figure}

Finally, in Fig.~\ref{fig:comparison_exp}(a) we show a scatter plot of the theoretical superconducting critical temperatures computed at three levels of theory (T$_{c}^{\rm AD}$, T$_{c}^{\rm iso}$, and T$_{c}^{\rm aniso}$) versus experimental results. 
Among our calculated database of 240 materials, we found 82 materials for which we could find an experimental T$_{c}$ with the corresponding crystal structure.
Note that for 31 of these 82 materials, we have not computed the T$_{c}^{\rm aniso}$ since the computed T$_{c}^{\rm iso}$ was lower than 5~K.
The data is scattered, although it is clear that the predicted T$_{c}$ overall overestimate experiment, which is well known in the literature~\cite{Boeri2022}. 

Interestingly, if we only take 51 materials for which we have T$_{c}^{\rm aniso}$ and collect data to remove materials with experimental mismatch with the simulated and the one with clear theoretical failure (listed above), we obtain a list of 37 materials shown in Fig.~\ref{fig:comparison_exp}(b).
We consider that curated set to be \emph{in reasonable agreement} with experimental data. 
However, we note that although T$_{c}^{\rm AD}$ may seem to be in closer agreement with the experiment in Fig.~\ref{fig:comparison_exp}(b), we expect to overestimate T$_{c}$ with first-principles calculations, since we are considering a perfect system and experimental samples often contain defects and imperfections, often reducing observed transition temperatures.
In that regard, T$_{c}^{\rm aniso}$ is more accurate and follows that expectation. 

%\newpage
%\clearpage
\section{Hubbard calculations}\label{sec:si-hubbard}

All the materials in our superconducting workflow have been computed to be non-magnetic at the standard DFT level. 
Additionally, for the 26 top candidates containing transition metals, we perform ground-state calculations using a Hubbard corrected density functional~\cite{Anisimov1991, Anisimov1995} to verify whether
a significant change of electronic structure occurs and whether a magnetic ground-state can be found with a +U correction. 
The Hubbard corrections can be used to reduce the self-interaction error of approximate semi-local functionals~\cite{Perdew:1981, MoriSanchez:2006}, and restore piece-wise linearity, which is particularly strong for elements having partially filled $d$ and $f$ orbitals.
In the following, we will refer to this kind of correction as DFT$+U^{\text{sc}}(+V^{\text{sc}})$, where $U^{\text{sc}}$ ($V^{\text{sc}}$) refers to the on-site (inter-site)\cite{Jr2010,Himmetoglu:2014} Hubbard parameter, and the subscript ``sc" stands for the self-consistent calculation of such parameter using constrained linear-response ~\cite{Cococcioni2005,Timrov:2021}. 
We compute the on-site parameters for any transition metal in the structure, and the inter-site parameter between such element and its first neighbors, if the neighbor belong to a reactive nonmetal (C, N, O, F, P, S, Cl, Se, Br, I).
The self-consistent procedure consists in an iterative loop of lattice vectors and atomic positions optimization at fixed $U^{\text{in}}~(V^{\text{in}})$, followed by a ground-state calculation which is used to perform the Hubbard parameters prediction, giving $U^{\text{out}}~(V^{\text{out}})$.
The latter output value is then used to start a new iteration until $U^{\text{in}}~(V^{\text{in}})$ and $U^{\text{out}}~(V^{\text{out}})$ are converged better than 0.1~eV for $U$ and 0.02~eV for $V$.
The Hubbard parameters are obtained from DFPT~\cite{Timrov:2018, Timrov:2021}, as implemented in the \textsc{HP} code~\cite{Timrov:2022} of the \textsc{Quantum ESPRESSO} suite.
The calculations are performed using a \textbf{k}-point distance of $0.2~\AA^{-1}$ for the ground-state calculation, and a \textbf{q}-point distance of $0.4~\AA^{-1}$ for DFPT.
The structural optimization is performed with a higher \textbf{k}-point sampling using a distance of $0.1~\AA^{-1}$.
We use orthogonalized atomic orbitals as Hubbard projector functions~\cite{Timrov:2020b} in all calculations.
The self-consistent cycle is carried out using the \texttt{aiida-quantumespresso-hp} package, which automates the entire procedure.

\begin{figure*}[h!]
    \centering
    \includegraphics[width=0.99\linewidth]{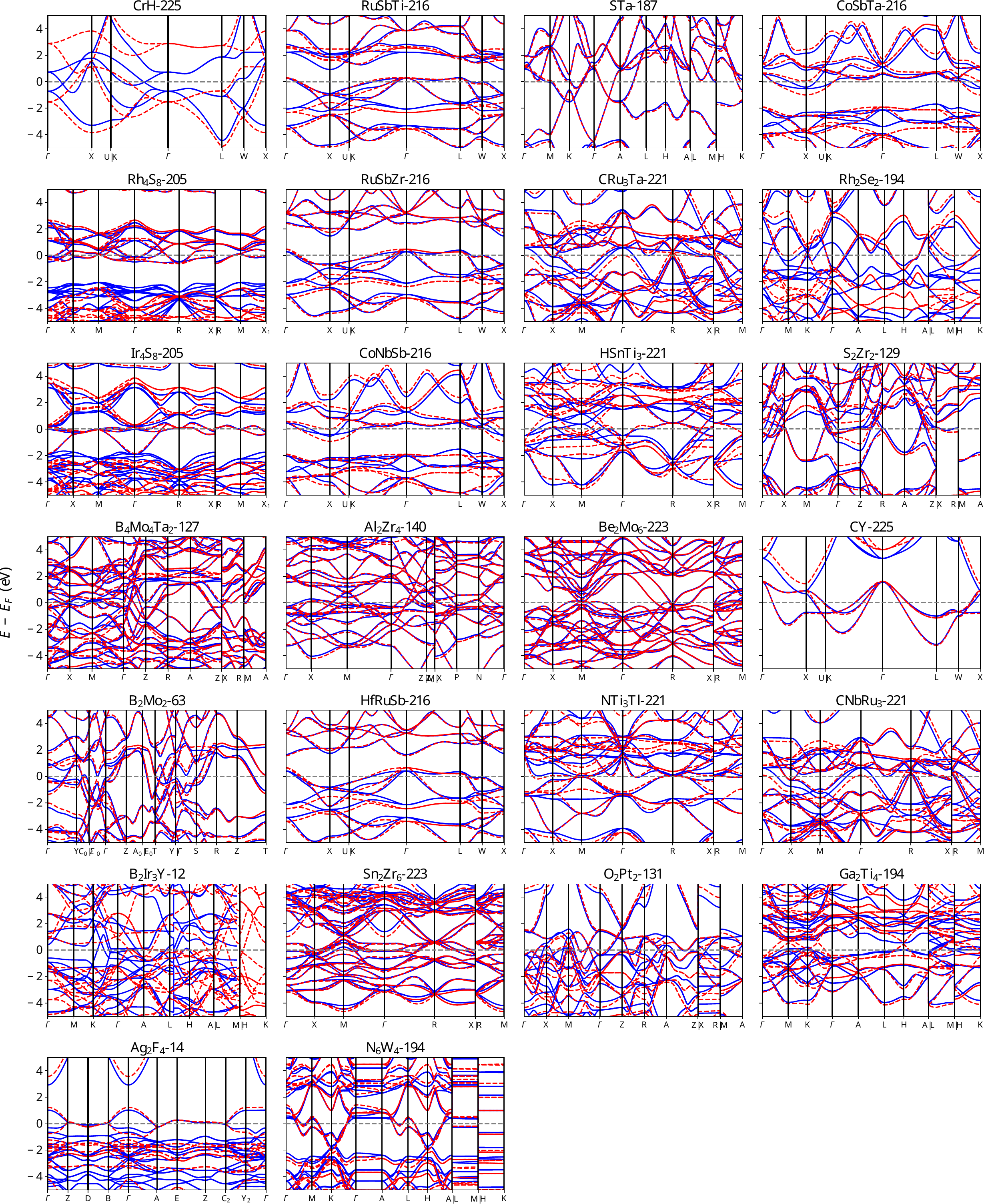}
    \caption{\label{fig:hubbard-bands}
    \textbf{Impact of $+U^{\text{sc}}(+V^{\text{sc}})$ Hubbard correction to the electronic bandstructure of potentially magnetic materials. }
     Comparison between PBE (blue) and PBE$+U^{\text{sc}}(+V^{\text{sc}})$ (dashed red) electronic band structure for the non-magnetic case. 
     The Hubbard parameters are obtained self-consistently and using density-functional perturbation theory.
     We note that in \ce{Rh2Se2} and \ce{YIr3B2}, the crystal symmetry reduces to a lower space group when optimized using PBE$+U^{\mathrm{sc}}(+V^{\text{sc}})$, explaining 
differences in high-symmetry path.  
    }
\end{figure*}

We first discuss the effect of Hubbard corrections on the electronic structure and magnetic configuration for the 26 top candidates containing transition metals.
For these materials, we start by calculating the self-consistent Hubbard parameters and ground-state for the non-magnetic (NM) case, using DFPT~\cite{Timrov2021}.
The resulting electronic band structures of PBE and PBE$+U^{\mathrm{sc}}(+V^{\mathrm{sc}})$ are compared in Figure~\ref{fig:hubbard-bands}.
We find that none of the materials undergoes a gap opening, confirming their metallic nature.
Except for the case of \ce{Rh2Se2} and \ce{B2Ir3Y}, the overall bandstructure are only weakly affected by the Hubbard corrections, especially close to the Fermi surface.

We now investigate the possibility of finding a magnetic ground-state when using Hubbard corrections.
We follow the same procedure as in the screening process and consider for simplicity the ferromagnetic (FM) configurations only, while saving a more in-depth search for the more interesting compounds. 
Since in constrained linear-response DFT the Hubbard parameters depend on the electronic response of the occupation matrices~\cite{Cococcioni2005}, we calculate the new set of self-consistent $U$ and $V$ parameters for the starting FM configurations, and compare the total energies with the corresponding NM partners.
The results are shown in Table~\ref{table:hubbard_nm_vs_m} where we find that 13 out of the 26 materials are potentially magnetic and we flag them as such in the tables of the main manuscript.

\begin{table*}[ht] 
\centering
\caption{\label{table:hubbard_nm_vs_m}
\textbf{Additional magnetic test for transition metal compounds.}
List of transition metal compounds where structurally self-consistent Hubbard $U$ and $V$ parameters are computed for non-magnetic (NM) and ferromagnetic (FM) configurations.
The total energies ($E^{\rm NM}$ and $E^{\rm FM}$) in eV units are reported as per-atom energy difference against the ground-state total energy ($0$ means ground-state).
The final absolute magnetization for the ferromagnetic calculations is report as $m^{\rm FM}$; a value of $0$ means that a local ferromagnetic ground-state is not found.
Self-consistent Hubbard $U$ and $V$ parameters are reported in eV units and are applied to the atom in red for $U$ and the red-blue atom pair for the $V$. 
When two $U$ are applied, their value is reported in the order they appear in the formula. 
A star (*) and a bold formatted magnetization indicate the materials that are ferromagnetic. 
}
\scriptsize
\begin{tabular}{llccccccccc}
\toprule
Formula & SG & $E^{\rm NM}$ & $E^{\rm FM}$ & $m^{\rm FM}$ & $U^{\rm NM}$ & $V^{\rm NM}$ & $U^{\rm FM}$ & $V^{\rm FM}$ \\
\hline
B$_2$\textcolor{red}{Mo$_{2}$}\textcolor{red}{Ta} & $P4/mbm$ & 0.000 & 0.000 & 0.0 & \makecell{3.7\\2.9} & - & \makecell{3.7\\2.9} & - \\
\textcolor{blue}{C}\textcolor{red}{Ru$_{3}$}\textcolor{red}{Ta}* & $Pm\overline{3}m$ & 0.085 & 0.000 & \textbf{2.9} & \makecell{4.9\\2.7} & \makecell{0.8--0.8} & \makecell{4.8\\2.7} & \makecell{0.8--0.8} \\
\textcolor{red}{Co}\textcolor{red}{Nb}Sb* & $F\overline{4}3m$ & 0.303 & 0.000 & \textbf{1.1} & \makecell{7.7\\2.8} & - & \makecell{6.4\\2.8} & - \\
\textcolor{red}{Co}Sb\textcolor{red}{Ta} & $F\overline{4}3m$ & 0.007 & 0.000 & 0.0 & \makecell{7.9\\2.5} & - & \makecell{7.9\\2.5} & - \\
\textcolor{red}{Cr}H* & $Fm\overline{3}m$ & 1.309 & 0.000 & \textbf{5.2} & \makecell{5.1} & - & \makecell{10.7} & - \\
HSn\textcolor{red}{Ti$_{3}$}* & $Pm\overline{3}m$ & 0.277 & 0.000 & \textbf{7.6} & \makecell{4.2} & - & \makecell{4.6} & - \\
\textcolor{red}{Ir}\textcolor{blue}{S$_{2}$}* & $Pa\overline{3}$ & 0.002 & 0.000 & \textbf{1.7} & \makecell{5.2} & \makecell{0.7--0.7} & \makecell{5.1} & \makecell{0.7--0.7} \\
\textcolor{red}{Rh}\textcolor{blue}{Se}* & $P6_3/mmc$ & 0.156 & 0.000 & \textbf{2.8} & \makecell{5.7} & \makecell{0.7--0.7} & \makecell{5.2} & \makecell{0.6--0.6} \\
\textcolor{red}{Rh}\textcolor{blue}{S$_{2}$}* & $Pa\overline{3}$ & 0.178 & 0.000 & \textbf{5.0} & \makecell{6.4} & \makecell{0.9--0.9} & \makecell{5.9} & \makecell{0.7--0.7} \\
\textcolor{red}{Ru}Sb\textcolor{red}{Ti}* & $F\overline{4}3m$ & 0.249 & 0.000 & \textbf{2.6} & \makecell{4.2\\4.1} & - & \makecell{4.1\\4.0} & - \\
\textcolor{red}{Ru}Sb\textcolor{red}{Zr} & $F\overline{4}3m$ & 0.000 & 0.000 & 0.0 & \makecell{4.3\\2.5} & - & \makecell{4.3\\2.5} & - \\
\textcolor{blue}{S}\textcolor{red}{Zr}* & $P4/nmm$ & 0.001 & 0.000 & \textbf{0.6} & \makecell{3.0} & \makecell{0.4--0.5} & \makecell{3.0} & \makecell{0.4--0.5} \\
\textcolor{blue}{S}\textcolor{red}{Ta} & $P\overline{6}m2$ & 0.000 & 0.000 & 0.0 & \makecell{3.0} & \makecell{0.4--0.4} & - & - \\
\textcolor{red}{Ag}\textcolor{blue}{F$_{2}$}* & $P2_1/c$ & 0.325 & 0.000 & \textbf{2.1} & \makecell{8.9} & \makecell{1.2--2.2} & \makecell{3.9} & \makecell{0.6--1.2} \\
Al\textcolor{red}{Zr$_{2}$} & $I4/mcm$ & 0.000 & 0.000 & 0.0 & \makecell{2.1} & - & \makecell{2.1} & - \\
B$_2$\textcolor{red}{Ir$_{3}$}\textcolor{red}{Y} & $C2/m$ & 0.025 & 0.000 & 0.0 & \makecell{1.9\\4.6} & - & \makecell{1.9\\4.6} & - \\
B\textcolor{red}{Mo} & $Cmcm$ & 0.000 & 0.000 & 0.0 & \makecell{3.8} & - & \makecell{3.8} & - \\
Be\textcolor{red}{Mo$_{3}$} & $Pm\overline{3}n$ & 0.000 & 0.000 & 0.0 & \makecell{3.1} & - & \makecell{3.1} & - \\
\textcolor{blue}{C}\textcolor{red}{Nb}\textcolor{red}{Ru$_{3}$}* & $Pm\overline{3}m$ & 0.233 & 0.000 & \textbf{4.3} & \makecell{3.0\\4.9} & \makecell{0.8--0.8} & \makecell{2.9\\4.5} & \makecell{0.8--0.8} \\
\textcolor{blue}{C}\textcolor{red}{Y} & $Fm\overline{3}m$ & 0.000 & 0.000 & 0.0 & \makecell{2.6} & \makecell{0.8--0.8} & \makecell{2.6} & \makecell{0.8--0.8} \\
Ga\textcolor{red}{Ti$_{2}$}* & $P6_3/mmc$ & 0.316 & 0.000 & \textbf{7.2} & \makecell{4.2, 4.2} & - & \makecell{4.2, 4.3} & - \\
\textcolor{red}{Hf}\textcolor{red}{Ru}Sb & $F\overline{4}3m$ & 0.000 & 0.000 & 0.0 & \makecell{2.1\\4.5} & - & \makecell{2.1\\4.5} & - \\
\textcolor{blue}{N$_{3}$}\textcolor{red}{W$_{2}$} & $P6_3/mmc$ & 0.000 & 0.000 & 0.0 & \makecell{3.9} & \makecell{0.7--1.1} & \makecell{3.9} & \makecell{0.7--1.1} \\
\textcolor{blue}{N}\textcolor{red}{Ti$_{3}$}Tl* & $Pm\overline{3}m$ & 0.683 & 0.000 & \textbf{5.0} & \makecell{4.7} & \makecell{0.8--0.8} & \makecell{4.3} & \makecell{0.7--0.7} \\
\textcolor{blue}{O}\textcolor{red}{Pt} & $P4_2/mmc$ & 0.000 & 0.000 & 0.0 & \makecell{4.7} & \makecell{0.7--0.7} & \makecell{4.7} & \makecell{0.7--0.7} \\
Sn\textcolor{red}{Zr$_{3}$} & $Pm\overline{3}n$ & 0.000 & 0.070 & 4.2 & \makecell{2.3} & - & \makecell{2.5} & - \\
\bottomrule
\end{tabular}
\end{table*}

While it has been shown that self-consistent Hubbard parameters provide accurate predictions of many materials properties~\cite{Timrov:2022}, one might still argue that total energies computed using different Hubbard parameters cannot be compared as one effectively changes the DFT functional.
To understand whether the relative energy difference between FM and NM changes when using the same $U$ and $V$ numerical values, we carry out two sets of single-point DFT calculations.
The first set uses self-consistent Hubbard parameters and optimized structures from the NM self-consistent calculations, and calculates single-point DFT ground-states for NM and FM configurations.
The second one does the same, but by using Hubbard parameters and structures from the FM self-consistent calculations.
We summarize in Table~\ref{table:hubbard_magnetic_check} the results of these calculations where we use ``x" to indicate when a material has a FM ground-state.
For the majority of the materials, the resulting ground-state is unchanged when using the same Hubbard parameters, either $U^{\mathrm{sc}}_{\mathrm{NM}}(V^{\mathrm{sc}}_{\mathrm{NM}})$ or $U^{\mathrm{sc}}_{\mathrm{FM}}(V^{\mathrm{sc}}_{\mathrm{FM}})$, or when compared using their self-consistent numerical values.

\begin{table}[ht] 
\centering
\captionsetup{width=0.95\linewidth}
\caption{
    \label{table:hubbard_magnetic_check}
    \textbf{Impact of self-consistency on the magnetic properties of transition metals materials.}
    List of transition metal compounds where non-magnetic (NM) and ferromagnetic (FM) total energy ground-states are compared using different Hubbard parameters and structures.
    We report an ``x" when the ground-state is found to be magnetic.
    The total energy differences are computed using Hubbard parameters and structure converged self-consistently (@SC), using structure obtained with a single shot ferromagnetic (@FM) calculations, and using a single shot non-magnetic (@NM) calculations.
}
\begin{tabular}{lccc}
\toprule
 Formula & @SC & @FM & @NM   \\
\hline
$\ce{Rh4S8}$ & $\times$ & $\times$ & $\times$ \\
$\ce{RuSbZr}$ &   &   &   \\
$\ce{Ag2F4}$ & $\times$ & $\times$ & $\times$ \\
$\ce{B2Ir3Y}$ &   &   &   \\
$\ce{B2Mo2}$ &   &   &   \\
$\ce{CoSbTa}$ &   &   &   \\
$\ce{NTi3Tl}$ & $\times$ & $\times$ & $\times$ \\
$\ce{CY}$ &   &   &   \\
$\ce{Al2Zr4}$ &   &   &   \\
$\ce{Be2Mo6}$ &   &   &   \\
$\ce{O2Pt2}$ &   &   &   \\
$\ce{N6W4}$ &   &   &   \\
$\ce{Rh2Se2}$ & $\times$ & $\times$ & $\times$ \\
$\ce{RuSbTi}$ & $\times$ & $\times$ & $\times$ \\
$\ce{B4Mo4Ta2}$ &   &   &   \\
$\ce{Ir4S8}$ & $\times$ & $\times$ & $\times$ \\
$\ce{CrH}$ & $\times$ & $\times$ & $\times$ \\
$\ce{HSnTi3}$ & $\times$ & $\times$ & $\times$ \\
$\ce{CRu3Ta}$ & $\times$ & $\times$ & $\times$ \\
$\ce{S2Zr2}$ & $\times$ & $\times$ & $\times$ \\
$\ce{CoNbSb}$ & $\times$ & $\times$ &   \\
$\ce{HfRuSb}$ &   &   &   \\
$\ce{CNbRu3}$ & $\times$ & $\times$ &   \\
\bottomrule
\end{tabular}
\end{table}

For the specific case of the four half-Heuslers \ce{NbCoSb}, \ce{TaCoSb}, \ce{TiRuSb} and \ce{ZrRuSb}, we study their magnetic states in more detail. 
The magnetic landscape is probed using a random forest search where the initial site-dependent magnetic configurations are constrained in a randomly initialized intersection of the found unique states, performed using the \textsc{ROMEO} code~\cite{Ponet2024}, and by applying the $+U_{\mathrm{FM}}$ correction reported in Table~\ref{table:hubbard_nm_vs_m}.
Since keeping the same on-site $U$ numerical values does not change the relative stability of the ground-states, it is therefore a legitimate approximation for our purposes. 
For completeness, we also perform the same exploration using PBE without any Hubbard corrections.
The magnetic search is performed in the conventional supercell (12 atoms) to account for more complex magnetic states. 
We perform the search by computing 100 random configurations per generation (\texttt{nrand} flag), constraining the Hubbard occupation for 100 self-consistent field steps (\texttt{Hubbard-maxstep} flag), and, after releasing the constraints, allowing up to 600 steps to converge the iterative DFT cycle (\texttt{electron-maxstep} flag). 
The algorithm is stopped when the ratio of newly discovered unique states to the number of trials is below 0.2 (\texttt{stopping-unique-ratio} flag) for 3 consecutive generations (\texttt{stopping-n-generations} flag). 
The \textbf{k}-point mesh for the DFT calculations is fixed for all compounds to a $11 \times 11 \times 11$ $\Gamma$-centered grid.

\begin{figure}[t] 
    \centering
    \includegraphics[width=0.99\linewidth]{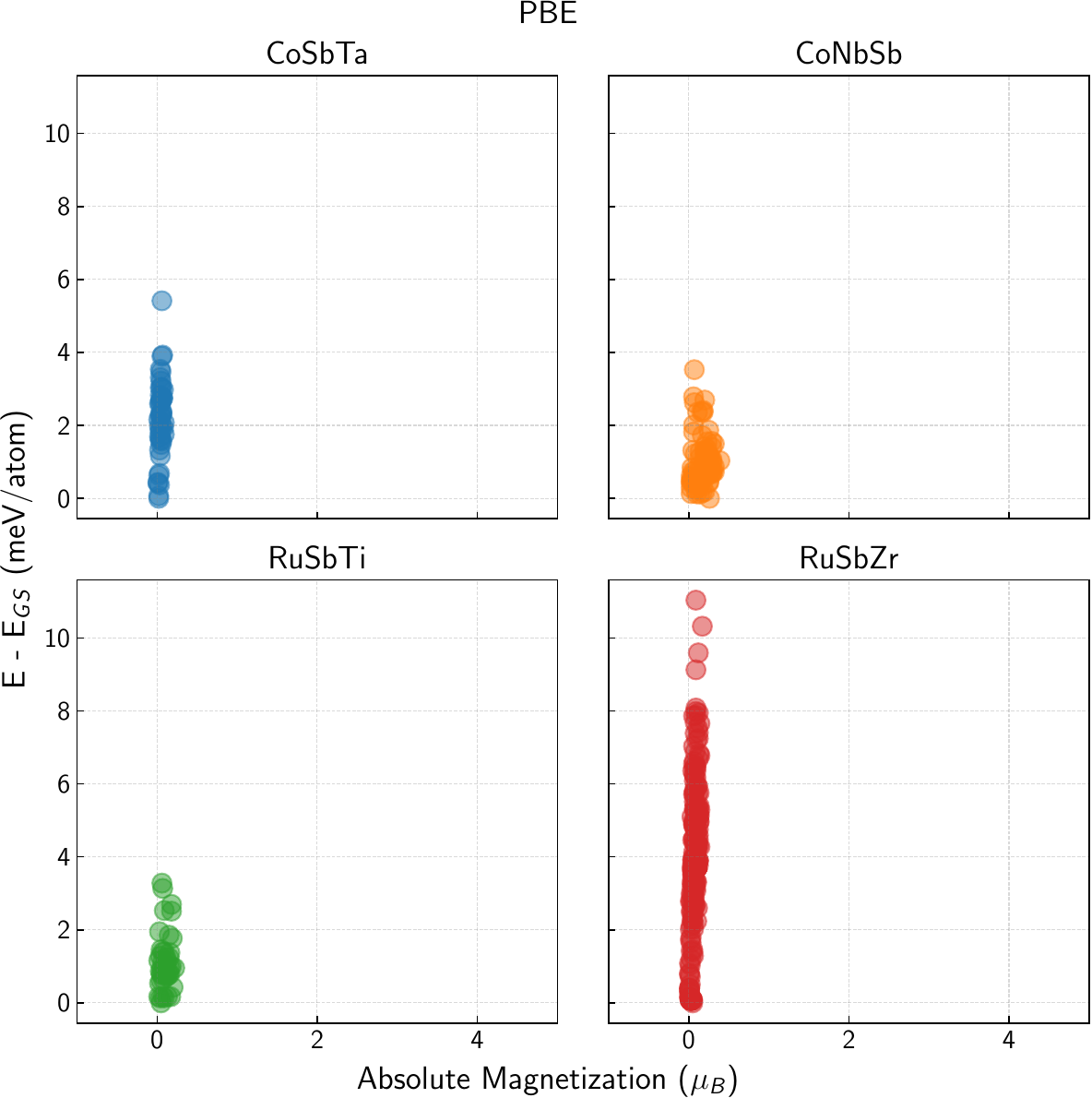}
    \caption{\label{fig:pbesearch}
    \textbf{Local magnetic ground state comparison using PBE. }
        Total energy difference as a function of the absolute magnetization of all the local magnetic ground-state found during the ROMEO search on the four half Heusler materials using the PBE functional.
    } 
\end{figure}

\begin{figure}[t]
    \centering
    \includegraphics[width=0.99\linewidth]{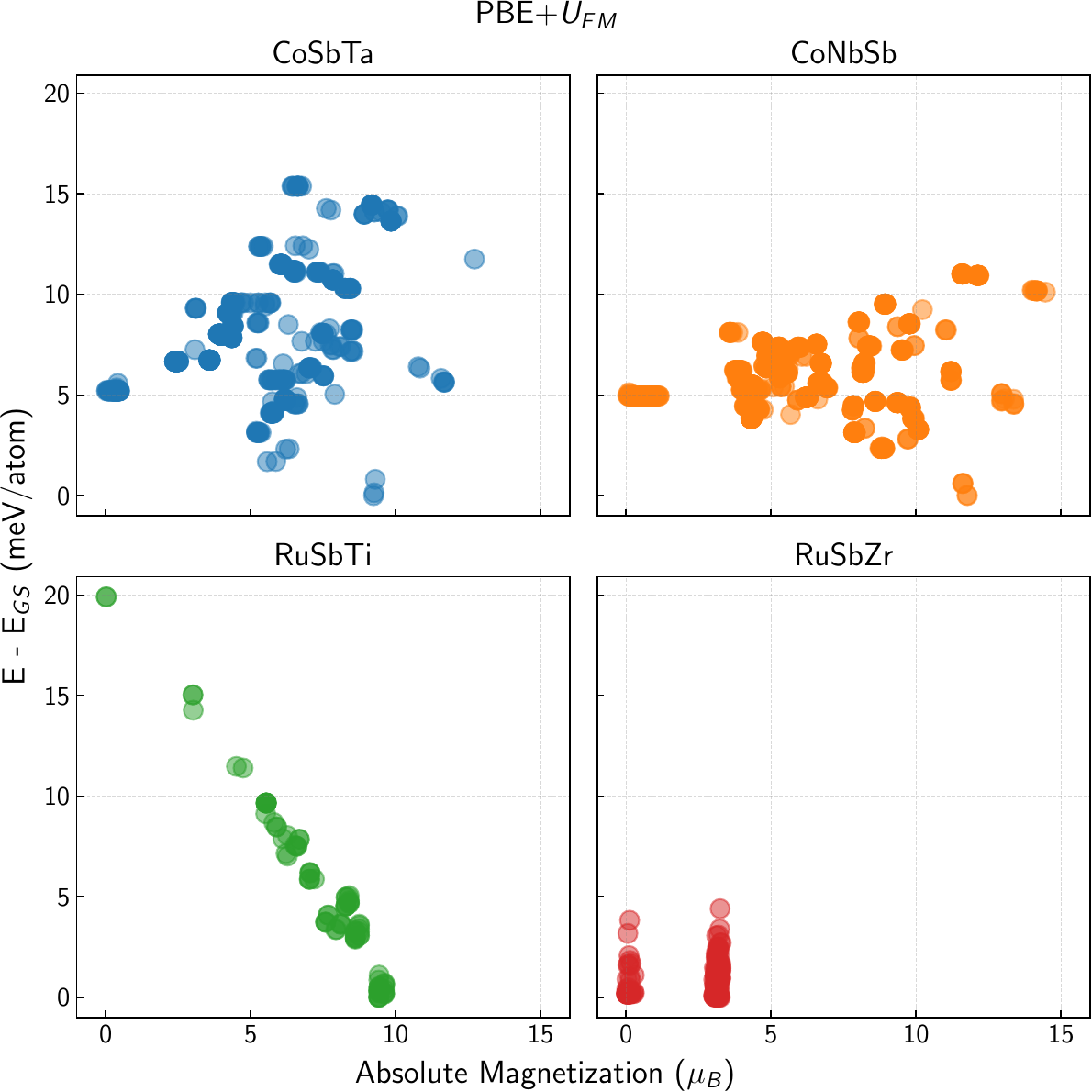}
    \caption{\label{fig:pbeusearch}
    \textbf{Local magnetic ground state comparison using PBE$+U$. }
        Total energy difference as a function of the absolute magnetization of all the local magnetic ground-state found during the ROMEO search on the four half Heusler materials using the PBE$+U$ functional, using the structures and $U$ parameters computed from the self-consistent Hubbard parameters calculation of the ferromagnetic (FM) configuration.
    }
\end{figure}

We show the magnetic energy landscape of each found configuration for the four half-Heuslers when using PBE in Fig.~\ref{fig:pbesearch} and PBE$+U_{\mathrm{FM}}$ in Fig.~\ref{fig:pbeusearch}.
In the case of PBE, we find that all configurations are non-magnetic but when using the $+U_{\mathrm{FM}}$, only the \ce{ZrRuSb} compounds is predicted to be non-magnetic.
Nevertheless, we note the small energy difference between the non-magnetic state and the magnetic ground-state, which is about 5~meV/atom for all four materials but for $\ce{RuSbTi}$, where the difference is about 20~meV/atom.

\clearpage

\section{Sensitivity analysis of the superconducting prediction}
% ---------------- SC TABLE ---------------- %
\setlength{\tabcolsep}{5pt}
\renewcommand{\arraystretch}{1.2}
\begin{table*}[t]
\centering
\caption{
\textbf{Impact of the Coulomb screening $\mu^*$ on superconducting properties of selected candidates.}
Comparison between experimental $T_c$ and calculated $T_c$ for part of the BCS superconductors with space Group and Pearson Symbol, alongside experimental and computed lattice parameters.}
\scriptsize
\begin{tabularx}{\textwidth}{lccccccccc}
\toprule
\textbf{Material} & \makecell{\textbf{Pearson} \\ \textbf{Symbol}} & \makecell{\textbf{Exp.}\\ \textbf{(a,c)}} & \makecell{\textbf{this work}\\ \textbf{(a,c)}}  & \multicolumn{3}{c}{\textbf{Allen-Dynes $T_c$ (K)}} & \makecell{\textbf{Iso.} \\ \textbf{$T_c$}} & \makecell{\textbf{Aniso.} \\ \textbf{$T_c$}} & \makecell{\textbf{Exp.} \\ \textbf{$T_c$}}\\
\cmidrule(lr){5-7}
 & & ($\AA$) & ($\AA$) & \textbf{$\mu^* \!\! = \! 0.1$} & \textbf{$\mu^* \!\! = \! 0.13$} & \textbf{$\mu^* \!\! =\! 0.2$} & (K) & (K) & (K) \\
\hline
\ce{Nb3Sn}-223 & cP8 & 5.29 & 5.32 & 19.40 & 18.30 & 15.70 & 25.60 & 35.58 & 18.05~\cite{Matthias1954a} \\
%\ce{MoN}-187 & hP2 & - & - & 30.10 & 26.70 & 19.10 & 32.10 & 36.60 & 12.00\cite{Matthias1952} \\
\ce{Tc}-194 & hP2 & 2.74, 4.40 & 2.75, 4.41 & 14.30 & 12.70 & 8.95 & 16.30 & 17.09 & 7.92~\cite{Giorgi_1966} \\
\ce{Nb}-229 & cI2 & 3.30 & 3.31 & 16.60 & 15.10 & 11.70 & 19.70 & 21.19 & 9.47~\cite{Boorse1960} \\
\ce{Pb}-225 & cF4 & 4.95 & 4.92 & 3.79 & 3.14 & 1.79 & 4.23 & - & 7.23~\cite{Pearson1958} \\
\ce{Ta3Sn}-223 & cP8 & 5.28 & 5.32 & 12.30 & 11.40 & 9.35 & 15.60 & 21.56 & 5.6~\cite{Wada_1973} \\
\ce{HfN}-225 & cF8 & 4.43 & 4.54 & 12.10 & 9.67 & 4.80 & 11.70 & 13.76 & 5.8~\cite{Potjan_2023} \\
\ce{V}-229 & cI2 & 3.03 & 3.00 & 22.20 & 20.40 & 16.20 & 27.08 & 29.09 & 5.13~\cite{Aaron1952} \\
\ce{In}-139 & tI2 & 4.60, 4.95 & 3.39, 4.79 & 2.04 & 1.93 & 1.68 & 5.13 & - & 3.41~\cite{Shaw1960, Bryant1961} \\
\ce{Re}-194 & hP2 & 2.76, 4.46 & 2.77, 4.49 & 7.06 & 5.62 & 2.81 & 7.69 & 8.30 & 2.42~\cite{Hulm1957, Daunt1952} \\ %
\ce{SbPt}-194 & hP4 & - & 4.23, 5.55 & 2.62 & 1.94 & 0.74 & 2.92 & - & 2.10~\cite{Matthias1953} \\
\ce{Te2Rh}-205 & cP12 & 6.44 & 6.45 & 3.82 & 3.07 & 1.62 & 4.37 & - & 1.51~\cite{Matthias1954c} \\
\ce{Ga}-63 & oC8 & 4.52, 7.66 & 4.53, 8.18 & 7.21 & 6.48 & 4.81 & 8.96 & 10.00 & 1.10~\cite{Seidel1958, Cochran1961} \\
\ce{TeRh}-194 & hP4 & 3.99, 5.66 & 4.11, 5.65 & 5.96 & 4.92 & 2.77 & 6.37 & - & 1.06~\cite{Matthias1954c} \\
\ce{TiNi}-221 & cP2 & 3.02 & 3.01 & 9.61 & 8.06 & 4.81 & 10.35 & 11.11 & 1.02~\cite{Matthias1963} \\
\ce{Nb3Sb}-223 & cP8 & 5.26 & 5.30 & 6.44 & 4.81 & 1.89 & 6.89 & - & 1.02~\cite{wood1958} \\
\ce{V3Sb}-223 & cP8 & 4.94 & 4.93 & 3.25 & 2.02 & 0.37 & 3.22 & - & 0.80~\cite{Matthias1963} \\
\ce{Zr}-194 & hP2 & 3.23, 5.15 & 3.23, 5.17 & 4.95 & 3.77 & 1.59 & 5.28 & - & 0.57~\cite{Smith1953, Smith1952} \\
\ce{Ti}-194 & hP2 & 2.95, 4.69 & 2.94, 4.65 & 5.78 & 4.25 & 1.58 & 6.09 & - & 0.49~\cite{Smith1953, Smith1952} \\
\bottomrule
\end{tabularx}
\label{table:superconductivity_comparison_extended_with_review_db}
\end{table*}
% ---------------- %
\balance
In Table.~\ref{table:superconductivity_comparison_extended_with_review_db}, we compare the Allen-Dynes $T_c^\text{AD}$ for three Coulomb screening $\mu^* = 0.10, 0.13, 0.20$, the isotropic Eliashberg $T_c^\text{iso.}$ and the anisotropoic Eliashberg $T_c^\text{aniso}$ with the experimental $T_c^\text{exp}$ for part of the BCS superconductors in our workchain.
The crystal structures are indicated by the combination of chemical formula and Pearson symbol and the experimental and computed lattice parameters are also reported. 
The table is ordered by the experimental $T_c^\text{exp}$.
We can see that the predictions can vary substantially between the different level of theories and that although the ordering is roughly preserved, there are some evident outliers and one should remain cautious.   

\begin{figure*}
    \centering
    \includegraphics[width=0.99\textwidth]{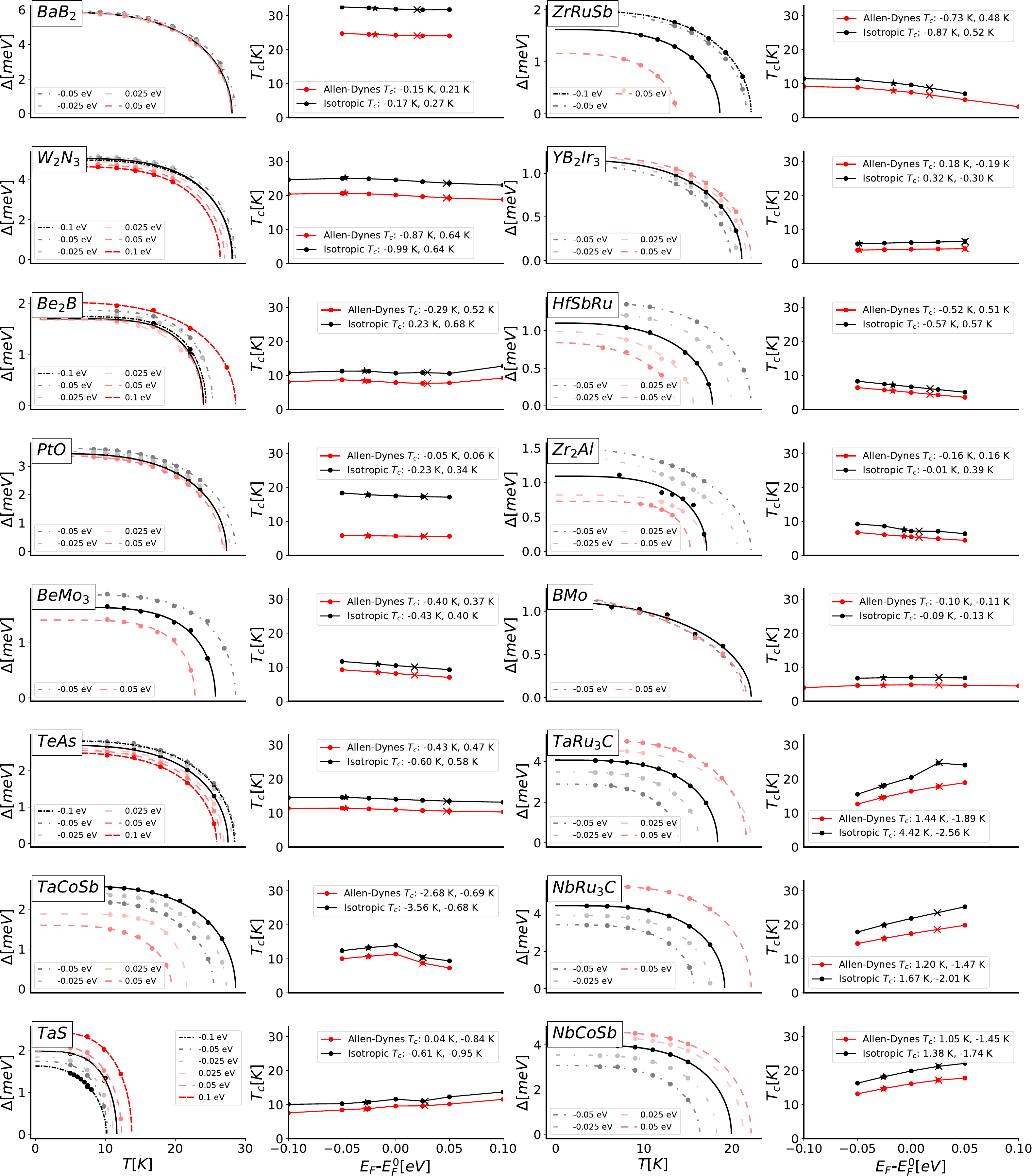}
    \caption{
    \textbf{Sensitivity analysis of the transition temperature versus Fermi level for the materials from Table II of the main manuscript (part 1).}
   Firstly the isotropic gap functions and fittings upon uniform dopings from -100~meV to 100~meV are shown on the left. 
   On the right side the extracted between Allen-Dynes $T_c$~(in red), isotropic $T_c$~(in black) and Fermi energy shift are plotted. 
   Critical electron/hole doping level~(10$^{21}$ cm$^{-3}$) are marked by cross/star and the data are marked by subscript and superscript in Table~(2) of the main manuscript.
    }
    \label{fig:sensitivity1}
\end{figure*}

\begin{figure*}
    \centering
    \includegraphics[width=0.99\textwidth]{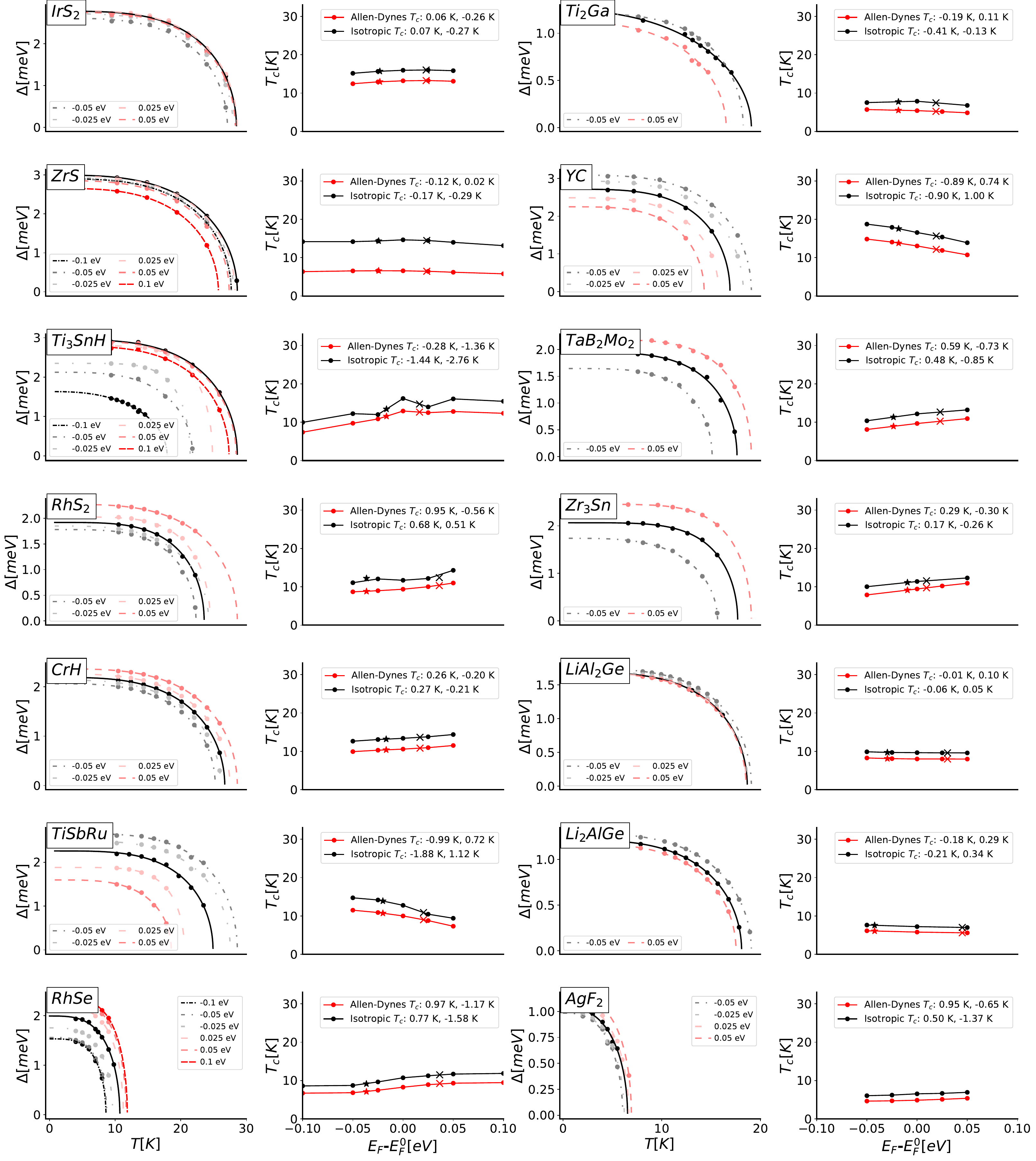}
    \caption{
    \textbf{Sensitivity analysis of the transition temperature versus Fermi level for the materials from Table II of the main manuscript (part 2).}
   Firstly the isotropic gap functions and fittings upon uniform dopings from -100~meV to 100~meV are shown on the left. 
   On the right side the extracted between Allen-Dynes $T_c$~(in red), isotropic $T_c$~(in black) and Fermi energy shift are plotted. 
   Critical electron/hole doping level~(10$^{21}$ cm$^{-3}$) are marked by cross/star and the data are marked by subscript and superscript in Table~(2) of the main manuscript.
    }
    \label{fig:sensitivity2}
\end{figure*}

To check the sensitivity of our prediction to errors in DFT, we check for all the top 30 unknown superconductors presented in Table~(2) of the main manuscript the sensitivity of \ADTC and \ISTC. 
As shown in Figs.~\ref{fig:sensitivity1} and \ref{fig:sensitivity2}, we firstly provide the fitting of \ISTC with different doping conditions, (fading black as hole doping, fading red as electron doping). 
Then we provide the change of \ADTC~(red) and \ISTC~(black) versus Fermi energy shifts. 

We consider 5 Fermi energy shifts, $\pm$0, $\pm$50~meV,$\pm$100~meV.
For materials with really sensitive \TC, we append another two shifts $\pm$ 25~meV. 
Finally, to correspond to the doping level used in experiment, we do self-consistent calculations where the electrons are added/removed by the amount of 10$^{21}$ cm$^{-3}$ as the simulation of homogeneously doping with electrons or holes. 
Then we find the corresponding Fermi energy shifts and we marked them on Figs.~\ref{fig:sensitivity1} and \ref{fig:sensitivity2} by cross~(for electron doping) and star~(for hole doping). 
Finally, superconductors whose absolute values of \TC variations are smaller than 0.5~K are viewed as \textit{robust} superconductors, and are highlighted in bold in Table~(2) of the main manuscript.

\newpage
\clearpage

\newpage

\clearpage

\section{Detailed analysis for \ce{BaB2}}\label{sec:BaB2}

\begin{figure*}[t]
    \centering
    \includegraphics[width=0.99\linewidth]{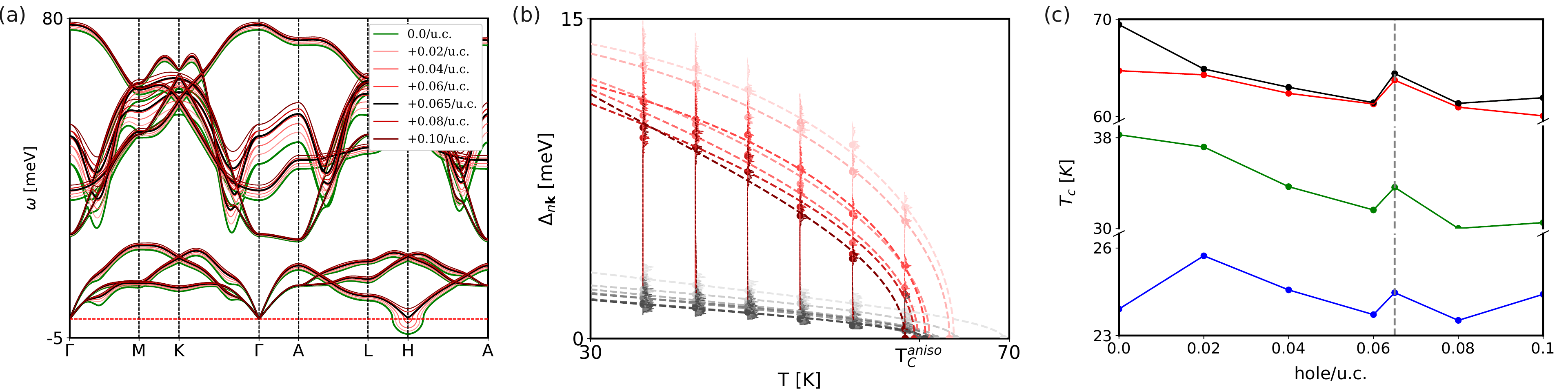}
    \caption{\label{fig:bab2_3}
    \textbf{Change of superconducting properties of BaB$_2$ with homogeneous hole doping.}
        Evolution of (a) phonon dispersions, (b) two anisotropic gaps, (c) Allen-Dynes \TC~(blue), isotropic \TC~(green), and anisotropic \TC~(red, black) at doping levels 0, 0.02, 0.04, 0.06, 0.08, 0.1 hole/unit cell. 
        The phonon dispersions of the undoped case and critical doping level 0.065 hole/unit cell (the doping used in the main manuscript) are emphasized in green and black, respectively.
        For this sensitivity analysis, we have used a smaller fine momentum grids of 40$\times$40$\times$24 $\mathbf{k}$ point and $\mathbf{q}$ point grids.
    }
\end{figure*}

\ce{BaB2} belongs to the \ce{MgB2} family of BCS superconductor and, like \ce{MgB2}, displays an unusual distinct two gap anisotropy.  
The relaxed pristine \ce{BaB2} has a soft mode at the $\mathbf{q}=\mathbf{H}$=($1/3, 1/3, 1/2$) point, as seen in Fig.~\ref{fig:bab2_3}(a) with a green line.  
We try to stabilize the structure by building a $3\times3\times2$ supercell to fold the $\mathbf{H}$ phonon back to $\Gamma$.
Then we impose random Gaussian displacements around 0.001\AA, on each atom and perform structural relaxation to find local minimum in the vicinity of the pristine phase. 
However, the stable lowest energy structure is forming clusters with high-energy phonon frequency that describe a different material.

Instead, we find that we can efficiently stabilize the structure with hole doping. 
As shown in Fig.~\ref{fig:bab2_3}(a), we apply several homogeneous hole doping and find that a hole doping of 0.065~hole per unit cell (1.46 $\cdot$ 10$^{21}$~cm$^{-3}$) is enough to stabilize the structure.
It is the value that we used in the main manuscript. 

As shown in Fig.~3c of the main manuscript, the lowest energy mode at the $\mathbf{H}$ point contributes to the electron-phonon coupling but only in a moderate way. 
Indeed, as shown in  Fig.~\ref{fig:bab2_3}(b-c), the superconducting gap and T$_{\rm c}$ are only weakly impacted by doping, changing by 5~K at most. 
Therefore, our prediction of high T$_{\rm c}$ in hole-doped \ce{BaB2} is robust and worth exploring experimentally. 
 
\begin{figure*}[t]
    \centering
    \includegraphics[width=0.99\linewidth]{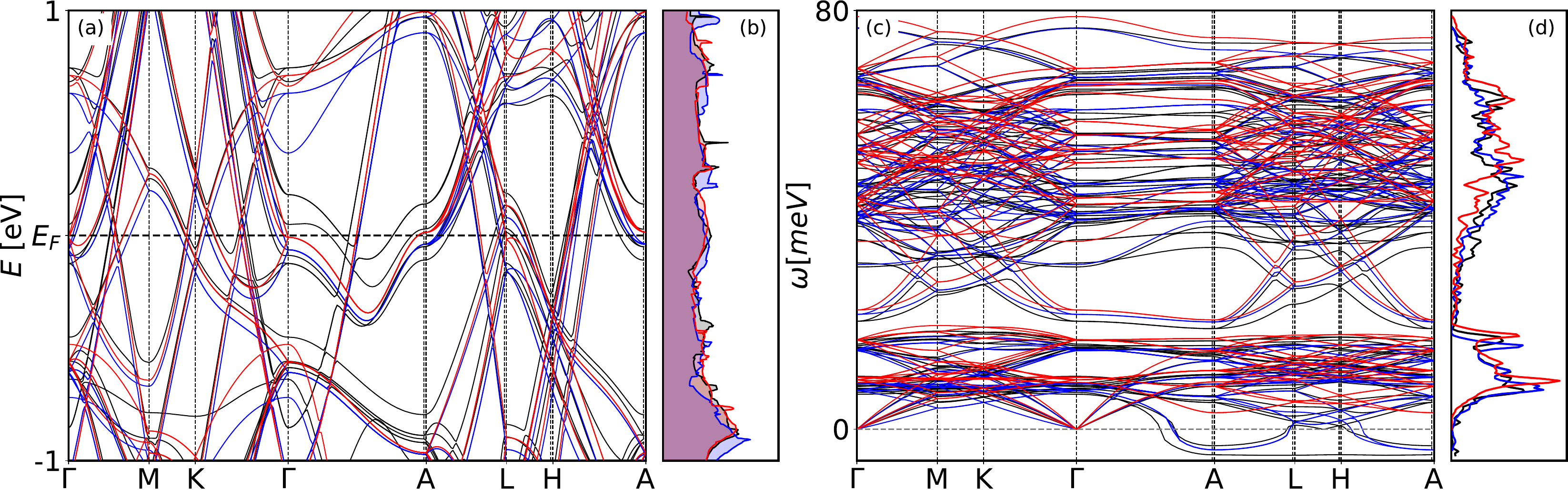}
    \caption{\label{fig:bab2}
    \textbf{Comparison between homogeneous hole doping and Cs-doping in BaB$_2$.}
        (a) Band structures, (b) electron density of states, (c) phonon dispersions and (d) phonon density of states of undoped $3\times3\times1$ supercell~(blue), 1 hole homogeneously doped in a $3\times3\times1$ supercell~(red) and 1 Cs-doped $3\times3\times1$ supercell~(black) \ce{BaB2}. In the electron band and dos, the energy are aligned to Fermi levels in each case.
    }
\end{figure*}

Finally, to verify that homogeneous doping is a good approximation, we computed the impact of Cs doping in a
$3\times3\times1$ supercell, corresponding to 0.111 hole doping.
In Fig.~\ref{fig:bab2} we compare the electron bandstructure and phonon bandstructure with their associated density of state for the case of undoped, homogeneously doped (0.111 hole/unit), and one Cs atom doped, all in a $3\times3\times1$ supercell.
In the undoped case, the instability folds back from the $\mathbf{H}$ to the $\mathbf{A}$=$(0, 0, 1/2)$ point, keeping the same soft mode value.
The homogeneous doping gives results that are consistent with Fig.~\ref{fig:bab2_3} and is stabilize. 
However, Cs doping is introducing local internal stresses that shift the Fermi energy up by 0.64~eV (see Fig.~\ref{fig:bab2}) preventing stabilization.  
We imposed random displacements and performed additional structural cell relaxation but could not find stabilization, pointing to the necessity to create an even larger supercell $3\times3\times2$.
However, such cell would lead to a 1/18 hole doping, which would not be enough. 
We would therefore need to simulate two Cs atoms in all possible configurations, which is out of the scope of this study.

\newpage

\onecolumn

\section{Full EPW pipeline results}\label{sec:si-epw-results}

Here we provide all the results for 250 superconductors with successful isotropic calculations, including (a) electron band structure from \textsc{Quantum ESPRESSO} (red) and \textsc{EPW} interpolation (black dash), (b) crystal structure, (c) phonon band structure, (d) spectral function $\alpha^2F(\omega)$ (black) and integrated spectral function $\lambda(\omega)$ (black dash).
For the successful 140 results of the anisotropic calculations, we also show (e) anisotropic gap functions (black) and their fittings.
The anisotropic $T_{\rm c}^{\rm aniso}$ is extracted from fitting the weighted average of the superconducting gap with the BCS gap equation. 
For the \textsc{EPW} parameters, FSR means solving the anisotropic Migdal-Eliashberg equation with Fermi-surface restricted integration~\cite{Lee2023} while FBW (IR) means full bandwidth using the intermediate representation~\cite{Mori2024} to speedup calculations.
Experimental T$_{\rm c}$ for known superconductors are provided for comparison with experimental references. 
We provide a ``\textbf{Note}" when there are potential issues between the source paper and the structure used or when the experimental T$_{\rm c}$ might be done on non-pristine materials. 
In such a case, the comparison between experimental and theoretical T$_{\rm c}$ must be done with care.  
We also report on the structural data used in the calculations and important computational parameters. 
All computational parameters used can be found in the \textsc{AiiDA} archive provided with this manuscript. 

\input{mpds-S1420799.tex} 
 
\newpage
 
\input{icsd-60614.tex}
\input{mpds-S1605022.tex}
\input{mpds-S1704431.tex}
\input{icsd-638417.tex}
\input{mpds-S378780.tex}
\input{mpds-S547970.tex}
\input{mpds-S1503730.tex}
\input{mpds-S555906.tex}
\input{mpds-S531232.tex}
\input{icsd-77216.tex}
\input{icsd-200101.tex}
\input{mpds-S529575.tex}
\input{mpds-S1937544.tex}
\input{mpds-S452438.tex}
\input{mpds-S542183.tex}
\input{mpds-S261186.tex}
\input{icsd-77392.tex}
\input{mpds-S1612207.tex}
\input{mpds-S1815445.tex}
\input{mpds-S1640543.tex}
\input{mpds-S453300.tex}
\input{mpds-S533193.tex}
\input{mpds-S1300517.tex}
\input{mpds-S457218.tex}
\input{mpds-S558472.tex}
\input{mpds-S451037.tex}
\input{mpds-S1825283.tex}
\input{mpds-S1929680.tex}
\input{mpds-S1210913.tex}
\input{mpds-S1637289.tex}
\input{mpds-S529740.tex}
\input{mpds-S554388.tex}
\input{icsd-659260.tex}
\input{mpds-S538753.tex}
\input{mpds-S554922.tex}
\input{mpds-S1120272.tex}
\input{mpds-S1612208.tex}
\input{mpds-S1605628.tex}
\input{mpds-S1708517.tex}
\input{mpds-S529571.tex}
\input{mpds-S1228295.tex}
\input{icsd-614887.tex}
\input{mpds-S313390.tex}
\input{mpds-S1612209.tex}
\input{icsd-192507.tex}
\input{mpds-S1828487.tex}
\input{icsd-614842.tex}
\input{mpds-S1627330.tex}
\input{mpds-S1708524.tex}
\input{mpds-S532171.tex}
\input{mpds-S460604.tex}
\input{mpds-S458288.tex}
\input{mpds-S1923735.tex}
\input{mpds-S1612210.tex}
\input{mpds-S551347.tex}
\input{cod-9012005.tex}
\input{mpds-S1815131.tex}
\input{mpds-S1530859.tex}
\input{mpds-S526947.tex}
\input{mpds-S536085.tex}
\input{mpds-S451405.tex}
\input{mpds-S1100285.tex}
\input{mpds-S1804230.tex}
\input{mpds-S549446.tex}
\input{mpds-S529452.tex}
\input{mpds-S525717.tex}
\input{mpds-S560356.tex}
\input{mpds-S261544.tex}
\input{mpds-S534295.tex}
\input{mpds-S1710219.tex}
\input{mpds-S1214986.tex}
\input{mpds-S534147.tex}
\input{icsd-43539.tex}
\input{mpds-S531588.tex}
\input{mpds-S1129391.tex}
\input{mpds-S300212.tex}
\input{mpds-S458122.tex}
\input{mpds-S558544.tex}
\input{mpds-S1626954.tex}
\input{icsd-56172.tex}
\input{mpds-S1927869.tex}
\input{mpds-S1404528.tex}
\input{mpds-S453173.tex}
\input{mpds-S1626947.tex}
\input{mpds-S1817003.tex}
\input{mpds-S532637.tex}
\input{mpds-S1932474.tex}
\input{mpds-S261398.tex}
\input{mpds-S1715746.tex}
\input{mpds-S379268.tex}
\input{mpds-S558779.tex}
\input{mpds-S1710218.tex}
\input{mpds-S457865.tex}
\input{mpds-S1500055.tex}
\input{mpds-S1404524.tex}
\input{mpds-S1240874.tex}
\input{mpds-S1708516.tex}
\input{mpds-S527728.tex}
\input{mpds-S451301.tex}
\input{mpds-S541753.tex}
\input{mpds-S459491.tex}
\input{mpds-S1833540.tex}
\input{mpds-S313403.tex}
\input{mpds-S460771.tex}
\input{mpds-S540352.tex}
\input{mpds-S1219456.tex}
\input{icsd-105937.tex}
\input{mpds-S1612211.tex}
\input{mpds-S528990.tex}
\input{mpds-S302655.tex}
\input{mpds-S530880.tex}
\input{mpds-S1628766.tex}
\input{mpds-S252001.tex}
\input{icsd-53646.tex}
\input{mpds-S260074.tex}
\input{mpds-S1020386.tex}
\input{mpds-S1921608.tex}
\input{icsd-639816.tex}
\input{mpds-S460112.tex}
\input{icsd-600055.tex}
\input{mpds-S251959.tex}
\input{mpds-S1251782.tex}
\input{mpds-S1714722.tex}
\input{mpds-S250521.tex}
\input{mpds-S1020948.tex}
\input{mpds-S1804665.tex}
\input{mpds-S459494.tex}
\input{mpds-S1613081.tex}
\input{mpds-S457333.tex}
\input{mpds-S1921345.tex}
\input{mpds-S1642209.tex}
\input{mpds-S1623893.tex}
\input{mpds-S551346.tex}
\input{icsd-35700.tex}
\input{mpds-S250560.tex}
\input{mpds-S1823490.tex}
\input{mpds-S553904.tex}
\input{icsd-617229.tex}
\input{mpds-S539785.tex}
\input{mpds-S378408.tex}
\input{mpds-S1500959.tex}
\input{mpds-S302110.tex}
\input{mpds-S1412199.tex}
\input{mpds-S551694.tex}
\input{mpds-S540350.tex}
\input{mpds-S553373.tex}
\input{mpds-S1830331.tex}
\input{mpds-S1816585.tex}
\input{mpds-S1627826.tex}
\input{mpds-S458045.tex}
\input{mpds-S553374.tex}
\input{mpds-S455725.tex}
\input{mpds-S1812974.tex}
\input{mpds-S261215.tex}
\input{icsd-105812.tex}
\input{mpds-S1005741.tex}
\input{mpds-S459493.tex}
\input{mpds-S1013663.tex}
\input{mpds-S1613078.tex}
\input{mpds-S1404523.tex}
\input{icsd-102016.tex}
\input{mpds-S2080133.tex}
\input{icsd-35701.tex}
\input{mpds-S1921607.tex}
\input{mpds-S451186.tex}
\input{mpds-S528319.tex}
\input{mpds-S535011.tex}
\input{mpds-S458244.tex}
\input{icsd-58794.tex}
\input{icsd-105158.tex}
\input{mpds-S301280.tex}
\input{mpds-S535628.tex}
\input{mpds-S528958.tex}
\input{mpds-S1253054.tex}
\input{mpds-S1800042.tex}
\input{icsd-57512.tex}
\input{mpds-S462063.tex}
\input{mpds-S1610171.tex}
\input{mpds-S260632.tex}
\input{mpds-S301111.tex}
\input{mpds-S452797.tex}
\input{mpds-S1822385.tex}
\input{mpds-S1833781.tex}
\input{mpds-S1014301.tex}
\input{mpds-S1012756.tex}
\input{mpds-S377145.tex}
\input{mpds-S1300855.tex}
\input{mpds-S532764.tex}
\input{mpds-S1234925.tex}
\input{mpds-S261126.tex}
\input{mpds-S1003472.tex}
\input{mpds-S1254092.tex}
\input{mpds-S531636.tex}
\input{mpds-S548863.tex}
\input{mpds-S1005744.tex}
\input{mpds-S311773.tex}
\input{mpds-S1323564.tex}
\input{mpds-S460718.tex}
\input{mpds-S312574.tex}
\input{mpds-S1822401.tex}
\input{icsd-58721.tex}
\input{icsd-77790.tex}
\input{mpds-S1709739.tex}
\input{mpds-S377144.tex}
\input{mpds-S527370.tex}
\input{mpds-S301153.tex}
\input{mpds-S534675.tex}
\input{mpds-S460717.tex}
\input{icsd-616428.tex}
\input{mpds-S1210907.tex}
\input{mpds-S1810286.tex}
\input{mpds-S1810194.tex}
\input{mpds-S1833544.tex}
\input{mpds-S1629062.tex}
\input{mpds-S532030.tex}
\input{cod-9016510.tex}
\input{mpds-S560756.tex}
\input{mpds-S526472.tex}
\input{mpds-S532862.tex}
\input{mpds-S528429.tex}
\input{mpds-S1014519.tex}
\input{mpds-S535758.tex}
\input{icsd-150970.tex}
\input{mpds-S560757.tex}
\input{mpds-S1801464.tex}
\input{mpds-S1641501.tex}
\input{mpds-S300058.tex}
\input{mpds-S377243.tex}
\input{cod-1538908.tex}
\input{mpds-S1229661.tex}
\input{icsd-74641.tex}
\input{mpds-S1630027.tex}
\input{mpds-S1927460.tex}
\input{mpds-S453106.tex}
\input{mpds-S527665.tex}
\input{mpds-S451339.tex}
\input{mpds-S453205.tex}
\input{mpds-S528833.tex}
\input{icsd-150969.tex}
\input{mpds-S1010614.tex}
\input{mpds-S452808.tex}
\input{mpds-S538754.tex}
\input{mpds-S1014455.tex}
\input{mpds-S1810752.tex}
\input{icsd-31084.tex}
\input{mpds-S1301076.tex}
\input{mpds-S300057.tex}
\input{mpds-S377147.tex}
\input{mpds-S1832004.tex}

\twocolumn